\documentclass[manyauthors,nocleardouble,COMPASS]{cernphprep}
\usepackage{amssymb}
\usepackage{amsmath}
\usepackage{amsbsy}
\usepackage{times}
\usepackage{epsfig}
\usepackage{colordvi}
\usepackage{graphicx}
\usepackage{wrapfig,rotating}
\usepackage{hyperref} 
\usepackage{multicol}
\usepackage{booktabs}
\usepackage{multirow}
\usepackage{lineno}

\RequirePackage{flushend}
\RequirePackage{placeins}
\RequirePackage{balance}
\RequirePackage[T1]{fontenc}

\makeatletter
\newsavebox\myboxA
\newsavebox\myboxB
\newlength\mylenA

\newcommand*\xoverline[2][0.6]{%
  \sbox{\myboxA}{$\m@th#2$}%
  \setbox\myboxB\null% Phantom box
  \ht\myboxB=\ht\myboxA%
  \dp\myboxB=\dp\myboxA%
  \wd\myboxB=#1\wd\myboxA% Scale phantom
  \sbox\myboxB{$\m@th\overline{\copy\myboxB}$}% Overlined phantom
  \setlength\mylenA{\the\wd\myboxA}%  calc width diff
  \addtolength\mylenA{-\the\wd\myboxB}%
  \ifdim\wd\myboxB<\wd\myboxA%
    \rlap{\hskip 0.7\mylenA\usebox\myboxB}{\usebox\myboxA}%
  \else
    \hskip -0.7\mylenA\rlap{\usebox\myboxA}{\hskip 0.5\mylenA\usebox\myboxB}%
  \fi}
\makeatother
\DeclareSymbolFont{letters}   {OML}{cmm}{m}{it}
\DeclareSymbolFont{symbols}   {OMS}{cmsy}{m}{n}
\DeclareSymbolFont{largesymbols}{OMX}{cmex}{m}{n}
\begin{document}
\begin{titlepage}
\PHnumber{2014--009}
\PHdate{January 23, 2014}

\title{Measurement of azimuthal hadron asymmetries in semi-inclusive deep inelastic 
scattering off unpolarised nucleons}

\Collaboration{The COMPASS collaboration}
\ShortAuthor{The COMPASS collaboration}

\begin{abstract}
Spin-averaged asymmetries in the azimuthal distributions of positive and
negative hadrons produced in deep inelastic scattering were measured using the
CERN SPS muon beam at $160$~GeV/c and a $^6$LiD target. The amplitudes of the
three azimuthal modulations $\cos\phi_h$, $\cos2\phi_h$ and $\sin\phi_h$ were
obtained binning the data separately in each of the relevant kinematic variables
$x$, $z$ or $p_T^{\,h}$ and binning in a three-dimensional grid of these three
variables. The amplitudes of the $\cos \phi_h$ and $\cos 2\phi_h$ modulations
show strong kinematic dependencies both for positive and negative hadrons.
\end{abstract}
\vfill
\Submitted{(to be submitted to Nucl.~Phys.~B)}
\end{titlepage}

{\pagestyle{empty}
%%%%%%%%%%%%%%%%%%%%%%%%%%%%%%%%%%%%%%%%%%%%%%%%%%%%%%%%%%%%%%%%%%%%%%%%%%%%%%%%%%%%%%%%%%%%%%%%%%%%%%%%%%%%%%%%%%%%%%%
%
% 2012_auththorlist.tex  
%
%%%%%%%%%%%%%%%%%%%%%%%%%%%%%%%%%%%%%%%%%%%%%%%%%%%%%%%%%%%%%%%%%%%%%%%%%%%%%%%%%%%%%%%%%%%%%%%%%%%%%%%%%%%%%%%%%%%%%%%

\section*{The COMPASS Collaboration}
\label{app:collab}
\renewcommand\labelenumi{\textsuperscript{\theenumi}~}
\renewcommand\theenumi{\arabic{enumi}}
\begin{flushleft}
C.~Adolph\Irefn{erlangen},
R.~Akhunzyanov\Irefn{dubna},
M.G.~Alekseev\Irefn{triest_i},
Yu.~Alexandrov\Irefn{moscowlpi}\Deceased,
G.D.~Alexeev\Irefn{dubna}, %1
A.~Amoroso\Irefnn{turin_u}{turin_i},
V.~Andrieux\Irefn{saclay},
V.~Anosov\Irefn{dubna}, %2
A.~Austregesilo\Irefnn{cern}{munichtu},
B.~Bade{\l}ek\Irefn{warsawu}
% ...
\end{flushleft}

%%%%%%%%%%%%%%%%%%%%%%%%%%%%%%%%%%%%%%%%%%%%%%%%%%%%%%%%%%%%%%%%%%%%%%%%%%%%%%%%%%%%%%%%%%%%%%%%%%%%%%%%%%%%%%%%%%%%%%%
%
% institutes
%
%%%%%%%%%%%%%%%%%%%%%%%%%%%%%%%%%%%%%%%%%%%%%%%%%%%%%%%%%%%%%%%%%%%%%%%%%%%%%%%%%%%%%%%%%%%%%%%%%%%%%%%%%%%%%%%%%%%%%%%

\begin{Authlist}
\item \Idef{bielefeld}{Universit\"at Bielefeld, Fakult\"at f\"ur Physik, 33501 Bielefeld, Germany\Arefs{f}}
\item \Idef{bochum}{Universit\"at Bochum, Institut f\"ur Experimentalphysik, 44780 Bochum, Germany\Arefs{f}\Arefs{ll}}
\item \Idef{bonniskp}{Universit\"at Bonn, Helmholtz-Institut f\"ur  Strahlen- und Kernphysik, 53115 Bonn, Germany\Arefs{f}}
\item \Idef{bonnpi}{Universit\"at Bonn, Physikalisches Institut, 53115 Bonn, Germany\Arefs{f}}
\item \Idef{brno}{Institute of Scientific Instruments, AS CR, 61264 Brno, Czech Republic\Arefs{g}}
\item \Idef{calcutta}{Matrivani Institute of Experimental Research \& Education, Calcutta-700 030, India\Arefs{h}}
\item \Idef{dubna}{Joint Institute for Nuclear Research, 141980 Dubna, Moscow region, Russia\Arefs{i}}
\item \Idef{erlangen}{Universit\"at Erlangen--N\"urnberg, Physikalisches Institut, 91054 Erlangen, Germany\Arefs{f}}
\item \Idef{freiburg}{Universit\"at Freiburg, Physikalisches Institut, 79104 Freiburg, Germany\Arefs{f}\Arefs{ll}}
\item \Idef{cern}{CERN, 1211 Geneva 23, Switzerland}
\item \Idef{liberec}{Technical University in Liberec, 46117 Liberec, Czech Republic\Arefs{g}}
\item \Idef{lisbon}{LIP, 1000-149 Lisbon, Portugal\Arefs{j}}
\item \Idef{mainz}{Universit\"at Mainz, Institut f\"ur Kernphysik, 55099 Mainz, Germany\Arefs{f}}
\item \Idef{miyazaki}{University of Miyazaki, Miyazaki 889-2192, Japan\Arefs{k}}
\item \Idef{moscowlpi}{Lebedev Physical Institute, 119991 Moscow, Russia}
\item \Idef{munichlmu}{Ludwig-Maximilians-Universit\"at M\"unchen, Department f\"ur Physik, 80799 Munich, Germany\Arefs{f}\Arefs{l}}
\item \Idef{munichtu}{Technische Universit\"at M\"unchen, Physik Department, 85748 Garching, Germany\Arefs{f}\Arefs{l}}
\item \Idef{nagoya}{Nagoya University, 464 Nagoya, Japan\Arefs{k}}
\item \Idef{praguecu}{Charles University in Prague, Faculty of Mathematics and Physics, 18000 Prague, Czech Republic\Arefs{g}}
\item \Idef{praguectu}{Czech Technical University in Prague, 16636 Prague, Czech Republic\Arefs{g}}
\item \Idef{protvino}{State Research Center of the Russian Federation, Institute for High Energy Physics, 142281 Protvino, Russia}
\item \Idef{saclay}{CEA IRFU/SPhN Saclay, 91191 Gif-sur-Yvette, France\Arefs{ll}}
\item \Idef{telaviv}{Tel Aviv University, School of Physics and Astronomy, 69978 Tel Aviv, Israel\Arefs{m}}
\item \Idef{triest_i}{Trieste Section of INFN, 34127 Trieste, Italy}
\item \Idef{triest_u}{University of Trieste, Department of Physics, 34127 Trieste, Italy}
\item \Idef{triest_ictp}{Abdus Salam ICTP, 34151 Trieste, Italy}
\item \Idef{turin_u}{University of Turin, Department of Physics, 10125 Turin, Italy}
\item \Idef{turin_i}{Torino Section of INFN, 10125 Turin, Italy}
\item \Idef{turin_p}{University of Eastern Piedmont, 15100 Alessandria, Italy}
\item \Idef{warsaw}{National Centre for Nuclear Research, 00-681 Warsaw, Poland\Arefs{n} }
\item \Idef{warsawu}{University of Warsaw, Faculty of Physics, 00-681 Warsaw, Poland\Arefs{n} }
\item \Idef{warsawtu}{Warsaw University of Technology, Institute of Radioelectronics, 00-665 Warsaw, Poland\Arefs{n} }
\item \Idef{yamagata}{Yamagata University, Yamagata, 992-8510 Japan\Arefs{k} }
\end{Authlist}
%%%%%%%%%%%%%%%%%%%%%%%%%%%%%%%%%%%%%%%%%%%%%%%%%%%%%%%%%%%%%%%%%%%%%%%%%%%%%%%%%%%%%%%%%%%%%%%%%%%%%%%%%%%%%%%%%%%%%%%
%
% Notes
%
%%%%%%%%%%%%%%%%%%%%%%%%%%%%%%%%%%%%%%%%%%%%%%%%%%%%%%%%%%%%%%%%%%%%%%%%%%%%%%%%%%%%%%%%%%%%%%%%%%%%%%%%%%%%%%%%%%%%%%%
\vspace*{-\baselineskip}\renewcommand\theenumi{\alph{enumi}}
\begin{Authlist}
\item \Adef{a}{Also at Instituto Superior T\'ecnico, Universidade de Lisboa, Lisbon, Portugal}
\item \Adef{aa}{Also at Department of Physics, Pusan National University, Busan 609-735, Republic of Korea and at Physics Department, Brookhaven National Laboratory, Upton, NY 11973, U.S.A. }
\item \Adef{bb}{Supported by the DFG Research Training Group Programme 1102  ``Physics at Hadron Accelerators''}
\item \Adef{b}{Also at Chubu University, Kasugai, Aichi, 487-8501 Japan\Arefs{k}}
\item \Adef{c}{Also at KEK, 1-1 Oho, Tsukuba, Ibaraki, 305-0801 Japan}
\item \Adef{cc}{Also at Moscow Institute of Physics and Technology, Moscow Region, 141700, Russia}
\item \Adef{y}{present address: National Science Foundation, 4201 Wilson Boulevard, Arlington, VA 22230, United States}
\item \Adef{x}{present address: RWTH Aachen University, III. Physikalisches Institut, 52056 Aachen, Germany}
\item \Adef{e}{Also at GSI mbH, Planckstr.\ 1, D-64291 Darmstadt, Germany}
\item \Adef{f}{Supported by the German Bundesministerium f\"ur Bildung und Forschung}
\item \Adef{g}{Supported by Czech Republic MEYS Grants ME492 and LA242}
\item \Adef{h}{Supported by SAIL (CSR), Govt.\ of India}
\item \Adef{i}{Supported by CERN-RFBR Grants 08-02-91009 and 12-02-91500}
\item \Adef{j}{\raggedright Supported by the Portuguese FCT - Funda\c{c}\~{a}o para a Ci\^{e}ncia e Tecnologia, COMPETE and QREN, Grants CERN/FP/109323/2009, CERN/FP/116376/2010 and CERN/FP/123600/2011}
\item \Adef{k}{Supported by the MEXT and the JSPS under the Grants No.18002006, No.20540299 and No.18540281; Daiko Foundation and Yamada Foundation}
\item \Adef{l}{Supported by the DFG cluster of excellence `Origin and Structure of the Universe' (www.universe-cluster.de)}
\item \Adef{ll}{Supported by EU FP7 (HadronPhysics3, Grant Agreement number 283286)}
\item \Adef{m}{Supported by the Israel Science Foundation, founded by the Israel Academy of Sciences and Humanities}
\item \Adef{n}{Supported by the Polish NCN Grant DEC-2011/01/M/ST2/02350}
\item [{\makebox[2mm][l]{\textsuperscript{*}}}] Deceased
\end{Authlist}

\clearpage}

\section{Introduction}
\label{intro}
\begin{figure}[ht]
\begin{center}
\includegraphics[width=.8\textwidth]{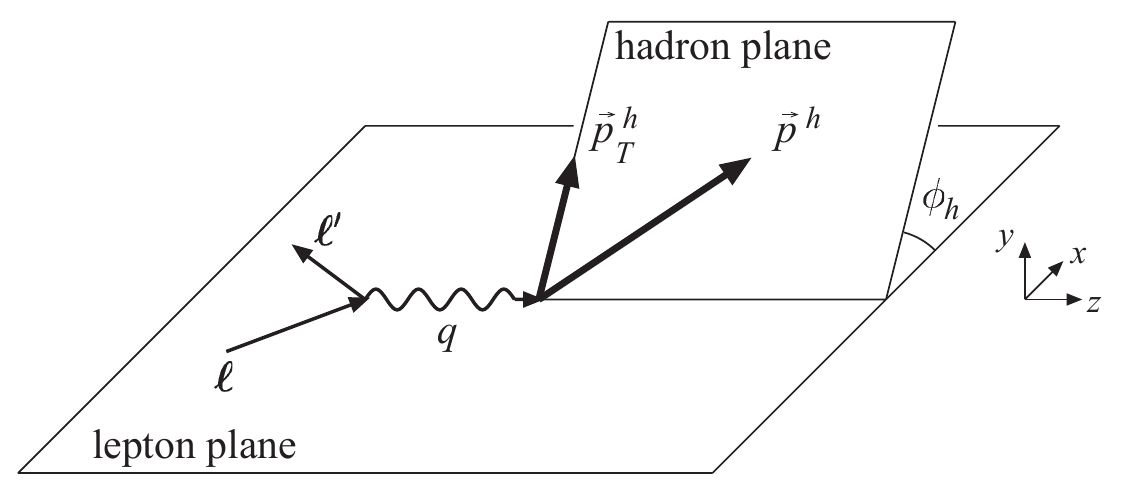}%
\end{center}
\caption{\label{fig:gns} SIDIS in the $\gamma^*N$ system: {$\vec{p}^{\,{h}}$}
is the momentum of the produced hadron and $\vec{p}^{\,{h}}_T$ its transverse 
component with respect to the virtual photon direction.}
\end{figure}

In the quark-parton model the transverse degrees of freedom of the nucleon
constituents are usually integrated over, and the parton distribution functions
(PDFs) as determined in lepton-nucleon deep inelastic scattering (DIS) depend
only on the Bjorken scaling variable $x$ and on $Q^2$, the virtuality of the
exchanged photon. On the other hand it was soon realised~\cite{feynman,ravndal}
that in semi-inclusive DIS (SIDIS) processes, namely in lepton-nucleon DIS in
which at least one hadron from the current jet is detected, a possible intrinsic
transverse momentum of the target quark would cause measurable effects in the
cross-section. Indeed the SIDIS cross-section is expected to exhibit a $\cos
\phi_h$ and a $\cos 2\phi_h$ modulation, where $\phi_h$ is the angle between the
lepton scattering plane and the plane defined by the hadron and the virtual
photon directions, as shown in Fig.~\ref{fig:gns}. The coefficients of these
modulations are predicted to vanish asymptotically as $1/Q$ and $1/Q^2$,
respectively~\cite{ravndal}. The early measurements in the 70s however were not
accurate enough to detect such modulations.

At the end of the 70s, interest in possible modulations of the SIDIS
cross-section came also from a different direction. Azimuthal asymmetries in
unpolarised processes in quantum chromodynamics (QCD) are generated by gluon
radiation and splitting, and the observation of these asymmetries was in fact
proposed as a test of perturbative QCD (pQCD)~\cite{georgi_pqcd}. Such a
possibility however was immediately questioned by R. Cahn~\cite{Cahn}. Using
simple kinematics the amplitudes of the azimuthal modulations expected from the
quark intrinsic transverse momentum could be computed and shown to be the
dominant term as long as both $Q^2$ and the hadron transverse momentum are not
too large~\cite{Cahn}. Azimuthal modulations in the SIDIS cross-section were
indeed first observed by the EMC Collaboration~\cite{EMC_1,EMC_2} and then at
FNAL~\cite{E665}, and at higher energies by the ZEUS experiment at
HERA~\cite{ZEUS}. The present understanding is that pQCD accounts for the
asymmetries at large values of the final-state hadron transverse momentum
$p_T^{\,h}$, while at low values ($p_T^{\,h} \lesssim 1$~GeV/$c$) it is the
intrinsic transverse motion of the quarks which plays the key
role~\cite{bar_smallpQCD}.

Intrinsic transverse momentum has recently attracted much attention in
connection with the great experimental and theoretical effort to understand the
origin of the nucleon spin and, in particular, the many transverse spin effects
in hadronic reactions observed since several decades. The PDFs of the nucleon
have been generalised to include this new degree of freedom, introducing the
transverse-momentum-dependent (TMD) distributions. Also, TMD fragmentation
functions (FF) have been introduced, the best known being the Collins FF, which
describes a correlation between the transverse momentum $\vec{p}_T^{\,h}$ of
each of the hadrons in a hadronic jet and the spin of the fragmenting quark in
the hadronization process of a transversely polarised quark. The knowledge of
this new sector of hadronic physics is still at its beginning, but several new
important phenomena have been assessed~\cite{bbm} within a solid theoretical QCD
framework~\cite{collins}. Within this framework, much attention has been payed
to distributions which are $T$-odd and for a long time were believed to be zero
to preserve $T$-invariance. It was demonstrated afterwards that either initial
or final state interactions can result in non-zero $T$-odd distributions. One
$T$-odd PDF, the Sivers function, has already been shown to be definitely
different from zero in SIDIS processes off transversely polarised protons, even
at high energies~\cite{hermes_siv,compass_siv}. Another $T$-odd TMD PDF is the
so-called Boer-Mulders (B-M) function, which describes the correlation between
the quark transverse spin and its transverse momentum in an unpolarised
nucleon~\cite{B-M}. On top of the Cahn effect, the B-M TMD PDF convoluted with
the Collins FF is expected to contribute to the amplitudes of the $\cos \phi_h$
and $\cos 2\phi_h$ modulations in unpolarised SIDIS processes and its extraction
from the cross-section data is an important goal of the more recent
investigations at lower energies by the HERMES Collaboration~\cite{hermes_unpol}
and by the CLAS Collaboration~\cite{clas}.

In this paper, first results on the azimuthal modulations in unpolarised SIDIS
obtained by the COMPASS experiment are presented. The paper is organised as
follows. Section~\ref{sec:sidis_xsection} summarises the formalism for the
SIDIS cross-section in the one-photon exchange approximation. A short
description of the experimental apparatus during the 2004 run is given in
Sect.~\ref{sec:compexp}. The data analysis, the method used to extract the
azimuthal asymmetries and the studies of the possible systematic effects are
described in Sections~\ref{sec:evSel},~\ref{sec:extraction} and ~\ref{sec:sys}.
Finally, the results are given in Sect.~\ref{sec:results}.

\section{The SIDIS cross-section}
\label{sec:sidis_xsection}
The spin-averaged differential SIDIS cross-section for the production of a
hadron $h$ with transverse momentum $p_T^{\,h}$ and a fraction $z$ of the
available energy is given in the one-photon exchange
approximation~\cite{bacchetta} by:
\begin{eqnarray}
\frac{\textrm{d}\sigma}{p_T^{\,h}\,\textrm{d}p_T^{\,h}\,\textrm{d}x\,\textrm{d}y\, \textrm{d}z\, \textrm{d}\phi_h} &=&
\sigma_0 \, \Big( 1 + \epsilon _1 A^{UU}_{\cos\phi_h} \cos \phi_h + 
\nonumber \\
&& + \epsilon _2 A^{UU}_{\cos 2\phi_h} \cos2\phi_h + \lambda \epsilon _3 A^{LU}_{\sin\phi_h} \sin\phi_h  \Big) \;,
\label{eq:cross_section_1}
\end{eqnarray}
where $\sigma_0$ is the $\phi_h$ independent part of the cross-section,
$\lambda$ is the longitudinal polarisation of the incident lepton, $y$ is the
fractional energy of the virtual photon, and the quantities $\epsilon _i$ are
given by:
\begin{eqnarray}
\epsilon _1 = \frac{2 (2-y)\sqrt{1-y} } {1+(1-y)^2 } , \; \;
\epsilon _2 = \frac{2 (1-y)} {1+(1-y)^2 } , \; \;
\epsilon _3 = \frac {2 y \sqrt{1-y} }{1+(1-y)^2} . 
\label{eq:epsilon}
\end{eqnarray}
The amplitudes $A^{XU}_{f(\phi_h)}$ will be referred to as azimuthal asymmetries
in the following. The superscripts \emph{UU} and \emph{LU} refer to unpolarised
beam and target, and to longitudinally polarised beam and unpolarised target,
respectively.

The $\cos\phi_h$ and the $\cos2\phi_h$ asymmetries are related to the Cahn
effect and to the B-M TMD PDF. The Cahn effect contributions to
$A^{UU}_{\cos\phi_h}$ and $A^{UU}_{\cos2\phi_h}$ originate from kinematics, when
the intrinsic transverse momenta $\vec{k}_T$ of quarks inside the nucleon is
taken into accouunt, starting from the elastic quark-lepton
cross-section~\cite{Cahn}. Also the B-M function contributes to both
$A^{UU}_{\cos\phi_h}$ and $A^{UU}_{\cos2\phi_h}$, where it appears convoluted
with the Collins FF. The $A^{LU}_{\sin\phi_h}$ asymmetry is due to higher-twist
effects and has no clear interpretation in terms of the parton model.

The amplitudes of the $\cos\phi_h$ and $\cos2\phi_h$ modulations have been
measured in SIDIS on unpolarised proton and deuteron targets in a kinematic
region similar to that of COMPASS by previous experiments~\cite{EMC_1,E665} and
at higher energies by the ZEUS experiment ~\cite{ZEUS}. Results at lower
energies have been recently published by HERMES~\cite{hermes_unpol} for positive
and negative hadrons separately and by CLAS~\cite{clas} for $\pi^+$.

COMPASS has presented preliminary results for $A^{UU}_{\cos\phi_h}$,
$A^{UU}_{\cos2\phi_h}$ and $A^{LU}_{\sin\phi_h}$ on the deuteron for positive
and negative hadrons in 2008~\cite{ferrara}. A more refined analysis on a 
limited phase space as well as the removal of some specific problems related to
the acceptance correction has lead the final results presented here. They have been
obtained from the data collected in 2004 with the transversely polarised $^6$LiD
target to measure the Collins and Sivers asymmetries~\cite{COMPASS_cs04}.

\section{The experimental apparatus}
\label{sec:compexp}

A brief description of the 2004 COMPASS apparatus is given in this Section.
More details on the COMPASS spectrometer can be found in
Ref.~\cite{COMPASS_spec}.

The $\mu^+$ beam was naturally polarised by the $\pi$ decay mechanism, and the
beam polarisation $\lambda$ was about $-80$\%. The beam intensity was
$2\cdot10^8$ $\mu^+$ per spill of 4.8~s with a cycle time of 16.8~s. The
$\mu^+$ momentum ($\sim 160$~GeV/$c$) was measured event by event in a Beam
Momentum Station (BMS) with a precision $\Delta p / p \lesssim 1\%$.

As the study of the nucleon spin was the main purpose of the experiment, a
polarised target system was used in 2004. It consisted of two cells, each 60~cm
long, filled with $^6$LiD, placed on the beam line, and housed in a cryostat
positioned along the axis of a solenoidal magnet. The $^6$LiD grains were
immersed in a mixture of liquid $^3$He~/~$^4$He. A small contamination of
$^7$Li almost exactly balances the proton excess in $^3$He, so that the target
can effectively be regarded to be isoscalar. The data used in the present
analysis (25\% of the full 2004 data sample) have been taken with the target
transversely polarised, i$.$e$.$ polarised along the direction of the dipole
field (0.42 T) provided by two additional saddle coils. The two target cells
were oppositely polarised, so data were taken simultaneously for the two target
polarization states. In order to keep systematic effects under control, the
orientation of the polarisation was reversed every 4 to 5 days (referred to as a
``period'' of data taking in the following). 

The spectrometer consists of two magnetic stages and comprises a variety of
tracking detectors, a RICH detector, two hadron calorimeters, and thick
absorbers providing muon identification. The first stage is centred around the
spectrometer magnet SM1, located 4~m downstream from the target centre, which
has a bending power of 1~Tm and a large opening angle to contain the hadrons of
the current jet. The second stage uses the spectrometer magnet SM2 (operated at
a bending power of 4.4 Tm), located 18~m downstream from the target, with an
acceptance of $\pm$50 and $\pm$25~mrad in the horizontal and vertical planes,
respectively. In order to match the expected particle flux at various locations
along the spectrometer, various tracking detectors are used. The small-area
trackers consist of several stations of scintillating fibres, silicon detectors,
micromegas chambers and gaseous chambers using the GEM technique. Large-area
tracking devices are made from gaseous detectors (Drift Chambers, Straw Tubes,
and MWPC's) placed around the two spectrometer magnets.

Muons are identified in large-area detectors using drift-tubes downstream of
iron or concrete absorbers. Hadrons are detected by two large iron-scintillator
sampling calorimeters, installed in front of the absorbers and shielded to avoid
electromagnetic contamination. The charged particle identification relies on
the RICH technology, but is not used in this analysis where results are given
for non-identified charged hadrons only.

In most DIS events the scattered muon is identified by coincidence signals in
the trigger hodoscopes which measure the particle trajectory in the vertical
(non-bending) plane and check its compatibility with the target position.
Several veto counters upstream of the target are used to avoid triggers due to
beam halo muons. In addition to this inclusive trigger mode, several
semi-inclusive triggers select events fulfilling requirements based on the muon
energy loss and on the presence of a hadron signal in the calorimeters. The
acceptance is further extended toward high $Q^2$ values by the addition of a
standalone calorimetric trigger in which no condition is set for the scattered
muon.

\section{Event selection and kinematic distributions}
\label{sec:evSel}
The DIS event and hadron selections are performed as in previous analyses based
on the same data~\cite{COMPASS_cs04}, and only a short description of the
procedure is given here.

A track reconstructed in the scintillating fibres and silicon detectors upstream
of the target is assumed to be an incoming muon if its momentum is measured in
the BMS. Scattered muons are selected among the positively charged outgoing
tracks with a momentum larger than $1$~GeV/$c$, passing through SM1. In order
to be accepted as the scattered muon, a track is required to cross an amount of
material in the spectrometer corresponding to at least $30$ radiation lengths
and must be compatible with the hits in the trigger hodoscopes. Only events
with one scattered muon candidate are accepted. The muon interaction point (the
so-called ``primary vertex'') is defined by one beam particle and the scattered
muon. The DIS events are selected requiring $Q^2>1$ (GeV/$c$)$^2$, $0.1<y<0.9$,
and an invariant mass of the hadronic final state system $W>5$ GeV/$c^2$.

If the amount of material traversed in the spectrometer is less than $10$
radiation lengths the outgoing particles are assumed to be hadrons. In order to
have a good resolution on the azimuthal angle the charged hadrons are required
to have at least $0.1$~GeV/$c$ transverse momentum $p_T^{\,h}$ with respect to
the virtual photon direction. In order to reject hadrons from target
fragmentation the hadrons are also required to carry a fraction $z>0.2$ of the
available energy while the contamination from hadrons produced in exclusive
reactions is reduced by requiring $z$ to be smaller than 0.85. No attempt is
made to further suppress diffractive meson production, as done e.g. in
Ref.~\cite{hermes_unpol}.

In addition to these standard requirements, further cuts have been applied
specific for this analysis because it requires acceptance corrected azimuthal
distributions of the final state hadrons. An upper limit on the transverse
hadron momentum has been introduced ($p_T ^{\,h} < 1.0$~GeV/$c$), both to ensure
negligible pQCD corrections and to obtain a better determined hadron acceptance.
In order to have a flat azimuthal acceptance the cut $\theta_{\gamma ^*} ^{~lab}
< 60$~mrad is applied, where $\theta_{\gamma ^*} ^{~lab}$ is the virtual photon
polar angle calculated with respect to the nominal beam direction in the
laboratory system. The cuts $y > 0.2$ and $x < 0.13$ have been also applied
because of the correlation of $x$ and $y$ with $\theta_{\gamma ^*} ^{lab}$.

The final event and hadron selection is thus:
\begin{itemize}
\item[] $Q^2 > 1$~(GeV/$c$)$^2$, \, $W > 5$~GeV/$c^2$, \,
$0.003 < x < 0.13$, \, $0.2 < y < 0.9$, 
\item[] $\theta_{\gamma ^*} ^{lab} < 60$~mrad, \, $0.2 < z < 0.85$ \,
and \, $0.1$~GeV/$c <p_T ^{\,h} < 1.0$~GeV/$c$. 
\end{itemize}
The statistics of the hadron sample after all cuts is given in
Table~\ref{tab:stat} for each of the 4 periods of data taken with the
transversely polarised $^6$LiD target in 2004. The data with opposite
polarisation have been combined after normalising them on the relative incoming
muon flux. The hadron standard sample consists mainly of
pions~\cite{compass_coll}, about 70\% for positive hadrons, 76\% in case of
negative hadrons. Positive kaons and protons amount to about 15\% each,
negative kaons and antiprotons amount to 16\% and 8\%, respectively, as
evaluated with a LEPTO Monte Carlo and cross-checked with the RICH detector.

\begin{figure}[tbh]
\begin{center}
\includegraphics[width=.45\textwidth]{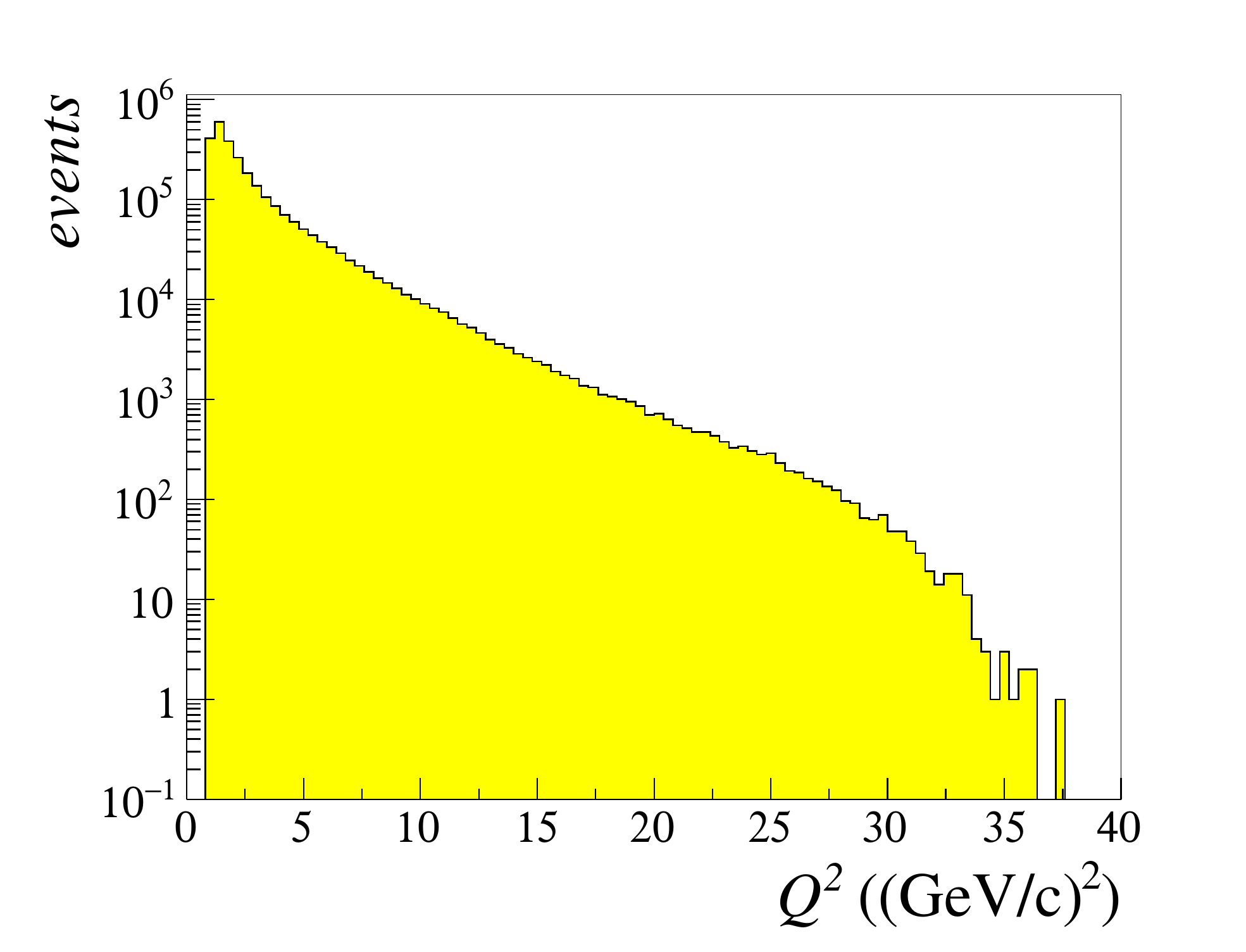} 
\includegraphics[width=.45\textwidth]{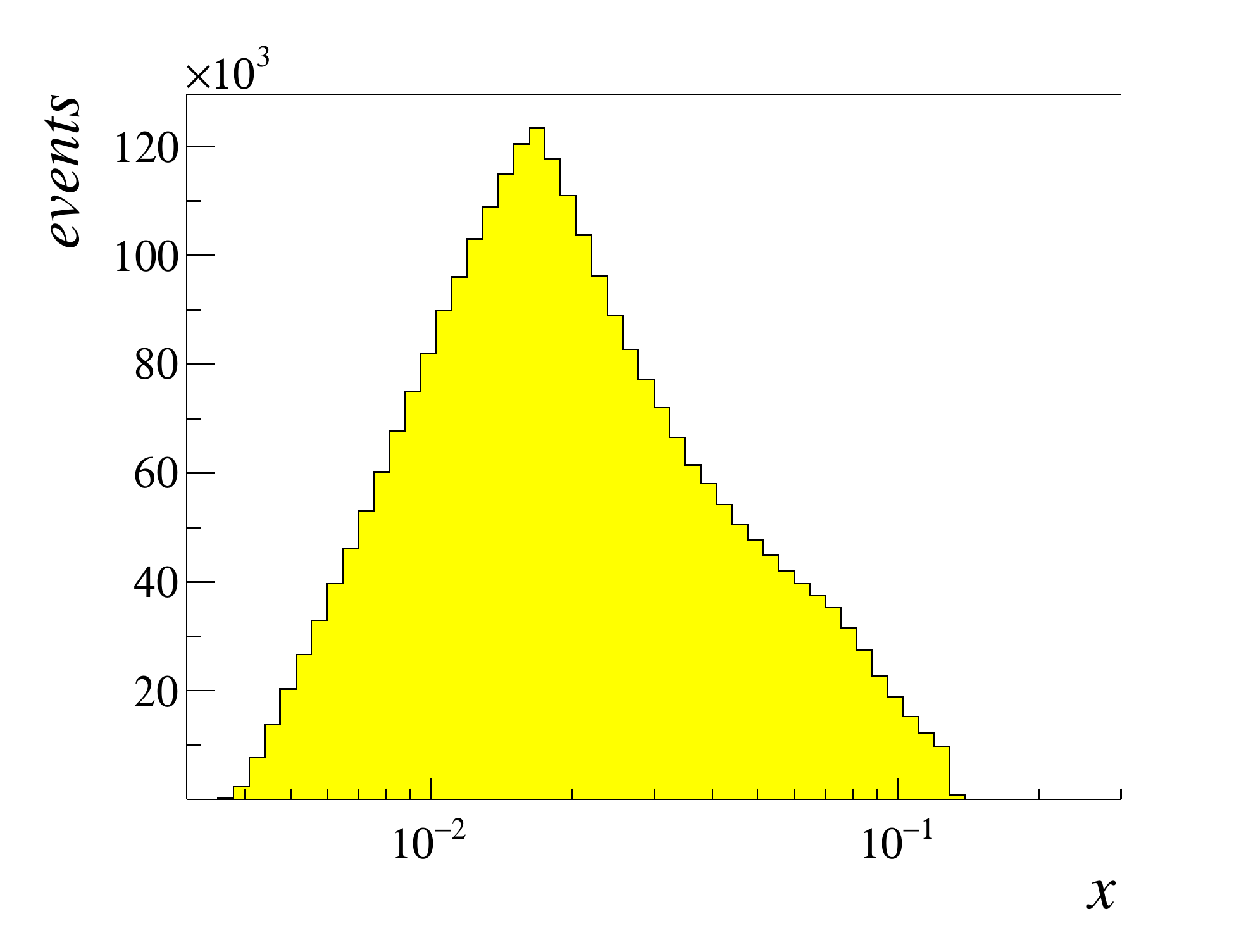} \hfill
\includegraphics[width=.45\textwidth]{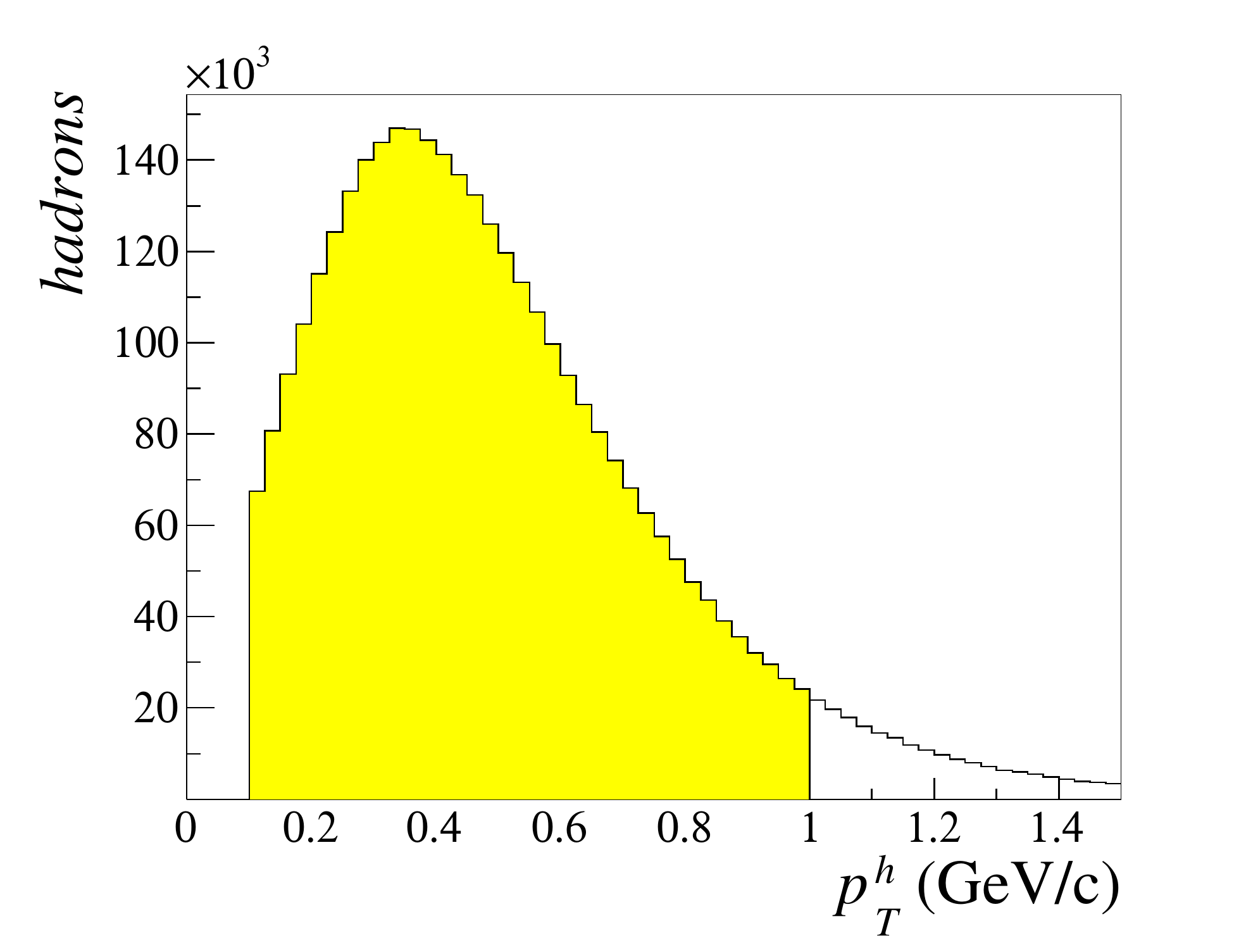} 
\includegraphics[width=.45\textwidth]{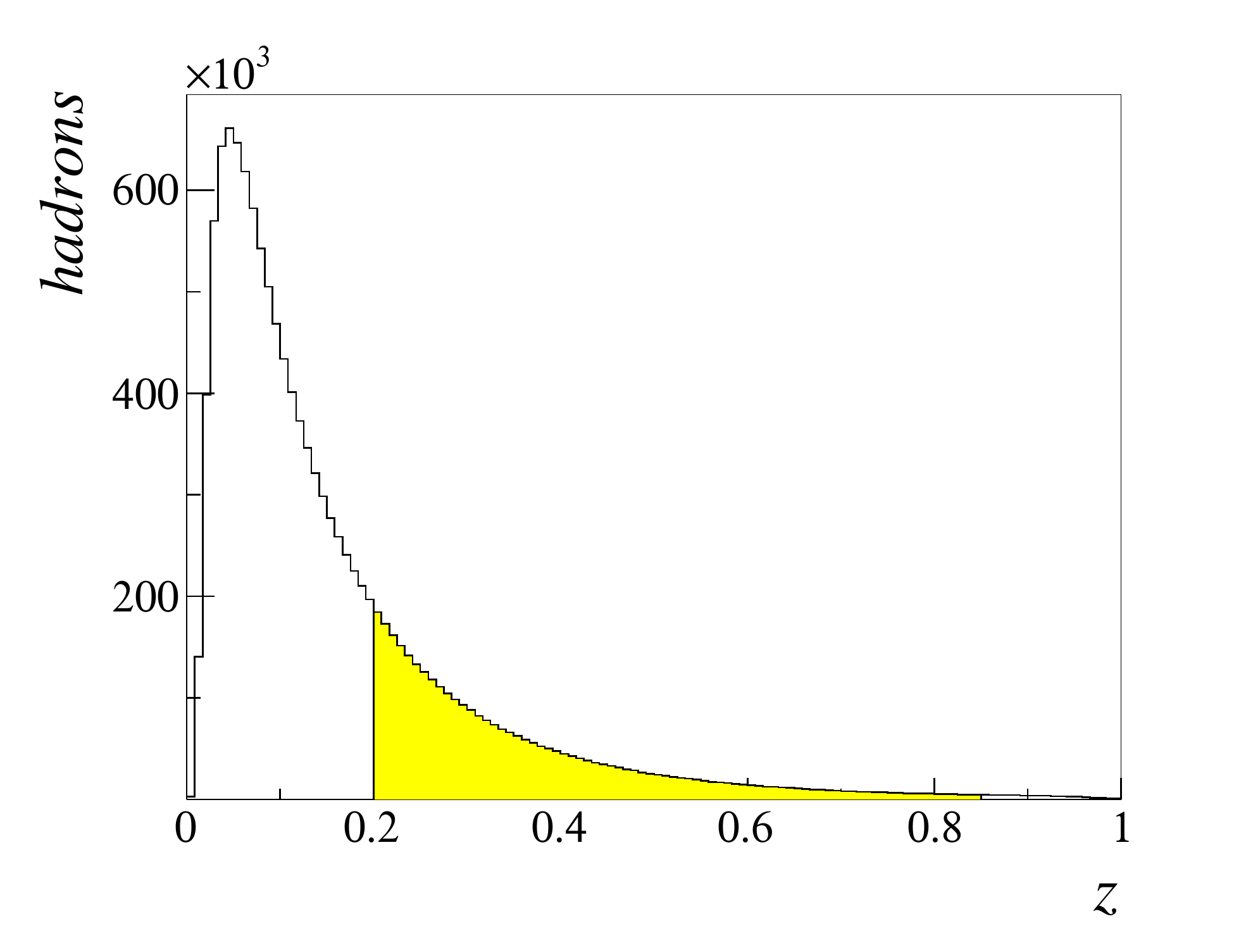} \hfill 
\end{center}
\caption{ 
Upper row: $Q^2$ and $x$ distributions of all the events in the final sample. 
Lower row: $p_T^{\,h}$ and $z$ hadron distributions for the same sample of events.}
\label{fig:dist}
\end{figure}

\begin{figure}[tbh]
\begin{center}
\includegraphics[width=.99\textwidth]{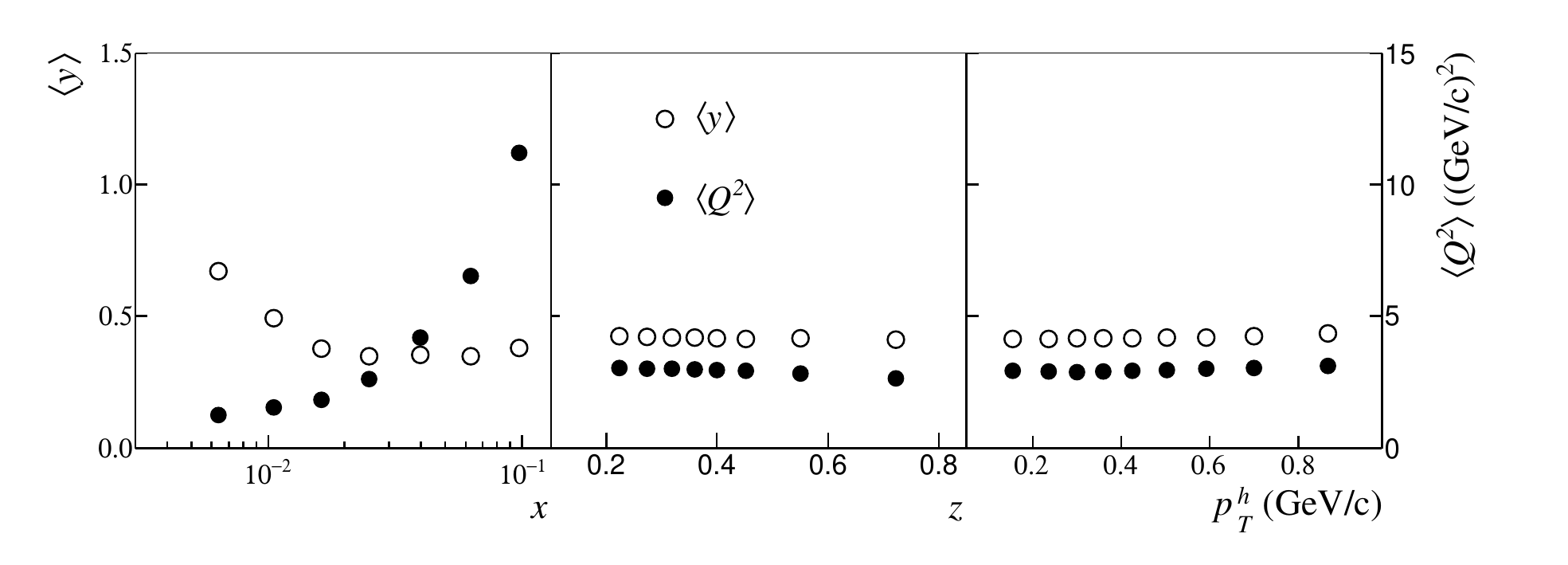} 
\end{center}
\caption{$Q^2$ and $y$ mean values calculated in the bins of $x$, of 
$z$ and of $p_T^{\,h}$.}
\label{fig:meanV}
\end{figure}

The $x$ distribution and the $Q^2$ distribution for the final sample are shown
in Fig.~\ref{fig:dist} together with the hadron $p_T^{\,h}$ and $z$
distributions. The mean values of $y$ and $Q^2$ with respect to $x$, $z$, and
$p_T ^{\,h}$ are shown in Fig.~\ref{fig:meanV}.
\begin{table*}[htb]
\caption{Final statistics used for the azimuthal asymmetry evaluation 
for each of the 4 data-taking periods.}
\begin{center}
\begin{tabular}{ c c c c}
\hline
period & positive hadrons & negative hadrons & polarisation\\
\hline
1 & 3.9$\, \cdot \, 10^5$ & 3.4$\, \cdot \, 10^5$ & +\\
2 & 3.4$\, \cdot \, 10^5$ & 2.9$\, \cdot \, 10^5$ & $-$\\
3 & 5.8$\, \cdot \, 10^5$ & 5.0$\, \cdot \, 10^5$ & +\\
4 & 3.6$\, \cdot \, 10^5$ & 3.1$\, \cdot \, 10^5$ & $-$\\
\hline
\end{tabular}%\\[2pt]
\end{center}
\label{tab:stat}
\end{table*}

\section{Extraction of the azimuthal asymmetries}
\label{sec:extraction}

\subsection{The method}
From Eq.~(\ref{eq:cross_section_1}), the measured azimuthal distributions are
expected to be:
\begin{eqnarray}
N(\phi_h, \vec{v}) &=& N_0(\vec{v}) \, a(\phi_h, \vec{v}) \, [ 1 + 
\epsilon_1 \, A^{UU}_{\cos\phi_h}(\vec{v})\, \cos\phi_h + \nonumber \\
&& + \epsilon_2 \, A^{UU}_{\cos2\phi_h}(\vec{v}) \, \cos2\phi_h 
+ \epsilon_3 \, \lambda \, A^{LU}_{\sin\phi_h}(\vec{v}) \, \sin\phi_h ], 
\label{eq:azi_distr}
\end{eqnarray}
where $a(\phi_h, \vec{v})$ is the apparatus acceptance and $\vec{v}$ indicates
the generic set of kinematic variables ($x$, $z$, $p_T^{\,h}$, ...) on which the
apparatus acceptance and the azimuthal asymmetries can depend. In order to
extract the azimuthal asymmetries it is necessary to correct the measured
azimuthal distributions by the $\phi_h$ dependent part of the apparatus
acceptance and to fit the corrected distribution with the appropriate $\phi_h$
modulation.

The azimuthal asymmetries have been first extracted from the data binned in $x$,
$z$ or $p_T^{\,h}$, and integrated over the other two variables (``integrated
asymmetries''). The bin widths have been chosen to be larger than the
experimental resolution estimated from Monte Carlo simulations. In each
kinematic bin the azimuthal distributions $N(\phi_h)$ are produced separately
for positive and negative hadrons, dividing the $(0,2\pi)$ $\phi_h$ range into
16 bins. The apparatus acceptance $a(\phi_h)$ is calculated from Monte Carlo
simulations for positive and negative hadrons for each bin of $\phi_h$ and for
each kinematic bin, as will be described in Sect.~\ref{sec:MC}. The hadron
azimuthal distributions corrected for the apparatus acceptance
$N_{corr}(\phi_h)\,=\, N(\phi_h) / a(\phi_h)$ are then fitted with a four
parameter function: $F(\phi_h)\,=\, p_0 \cdot (1\,+\,p_{\cos\phi_h}\cdot
\cos\phi_h\,+\,p_{\cos2 \phi_h} \cdot \cos2 \phi_h\,+\,p_{\sin\phi_h}\cdot
\sin\phi_h)$. The azimuthal asymmetries are then obtained by dividing the
fitted parameters by the appropriate quantities, i.e.:
\begin{equation}
A^{UU}_{\cos\phi_h} = \frac{p_{\cos\phi_h}}{\langle \epsilon_1 \rangle}, \,\, 
A^{UU}_{\cos2\phi_h} = \frac{p_{\cos2\phi_h}}{\langle\epsilon_2 \rangle}, \,\, 
A^{LU}_{\sin\phi_h} = \frac{p_{\sin\phi_h}}{\langle \epsilon_3 \rangle \lambda}.
\end{equation}
The quantities $\langle \epsilon_i \rangle$ are the mean values of $\epsilon_i$
defined in Eq.~(\ref{eq:epsilon}) and calculated for each kinematic bin. The
two central bins in $\phi_h$ have been excluded from the fit as will be
explained in Sect.~\ref{sec:rad}.

The same procedure is used to measure the azimuthal asymmetries for the hadrons
binned simultaneously in $x$, $z$ and $p_T^{\,h}$ (``3d asymmetries'').

\subsection{Monte Carlo and acceptance corrections}
\label{sec:MC}

In each kinematic bin and for each $\phi_h$ bin the azimuthal acceptance has
been evaluated as:
\begin{equation}
\label{eq:acc1}
a(\phi_{h_i}) \,=\, N_{rec}(\phi_{h_i}) / N_{gen}(\phi_{h_i}),
\end{equation}
where $N_{rec}(\phi_{h_i})$ is the number of reconstructed hadrons obtained from
the Monte Carlo simulation and $N_{gen}(\phi_{h_i})$ is the corresponding number
of generated hadrons. In order to obtain the number of reconstructed hadrons
the same kinematic cuts, the same event reconstruction, and the same event and
hadron selection as for the real data have been applied. Only the kinematic
cuts are applied to evaluate the number of generated hadrons.

The simulation involves the full COMPASS Monte Carlo chain, namely: the
generation of the DIS event, the propagation of the event inside the apparatus,
and the reconstruction of particle tracks. The LEPTO generator~\cite{lepto} is
used for the first step. The interactions between particles and materials and
the detectors response are simulated using COMGEANT, a software based on
GEANT3~\cite{geant} and developed inside the Collaboration to describe the
COMPASS set-up and which also includes trigger efficiencies,while detector
efficiencies are simulated at CORAL level. The
package CORAL~\cite{coral} is used to perform the track reconstruction and it is
the same program used for the real data. It has been carefully checked that the
Monte Carlo simulation gives a good description of the apparatus.

Starting from the distributions obtained using the default LEPTO setting,
different tunings of the LEPTO parameters and also different sets of PDFs,
already tested in other COMPASS analysis~\cite{highPT}, have been used. The
CTEQ5~\cite{cteq5} PDF set and the tuning of Ref.~\cite{highPT} have been
adopted for the extraction of the acceptances.

\begin{figure}[tbh]
\begin{center}
\includegraphics[width=.8\textwidth]{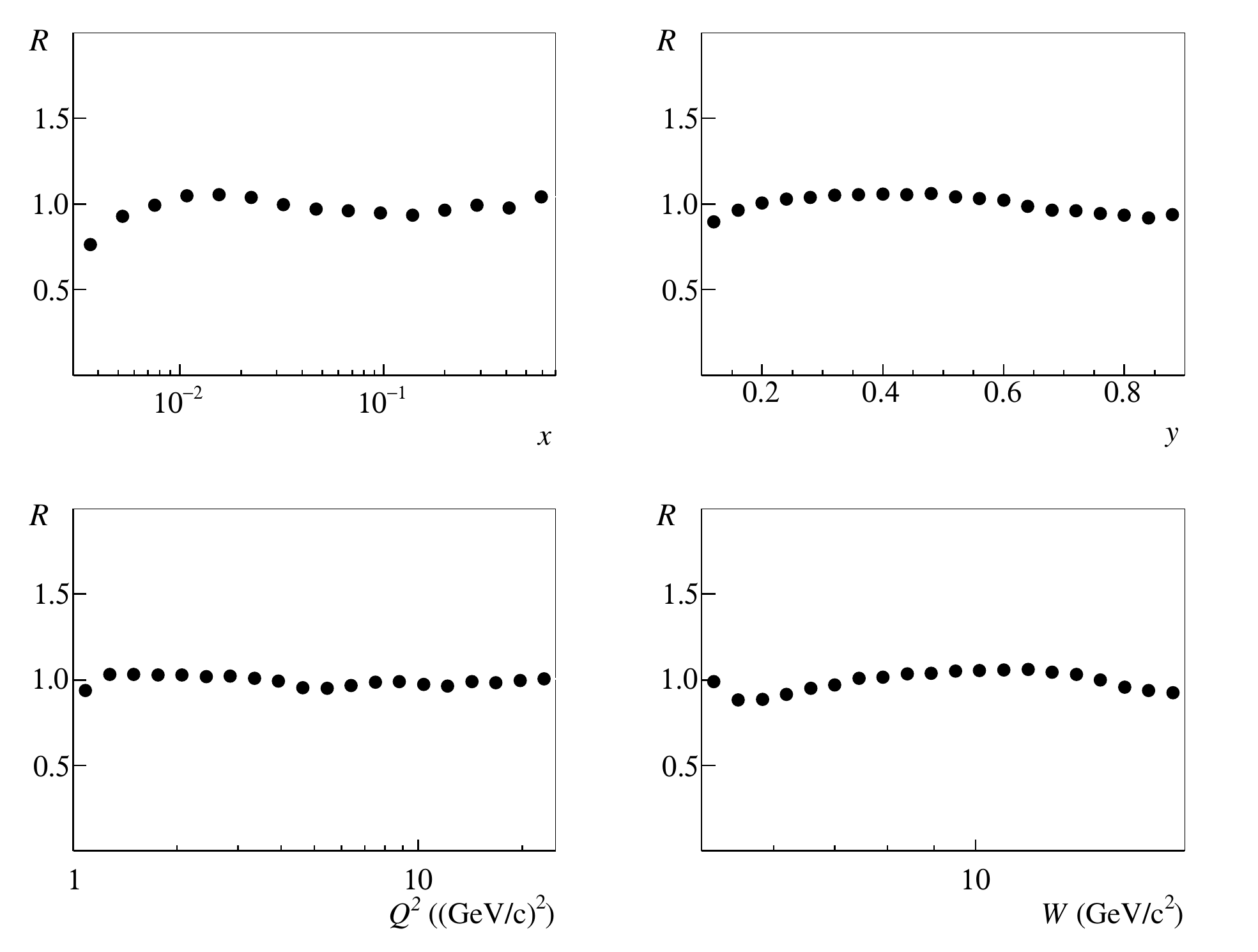} 
\end{center}
\caption{ Ratio $R$ between data and Monte Carlo events distributions 
for $x$, $y$, $Q^2$ and 
$W$. }
\label{fig:mc_rd_cmp1}
\end{figure}

\begin{figure}[tbh]
\begin{center}
\includegraphics[width=.8\textwidth]{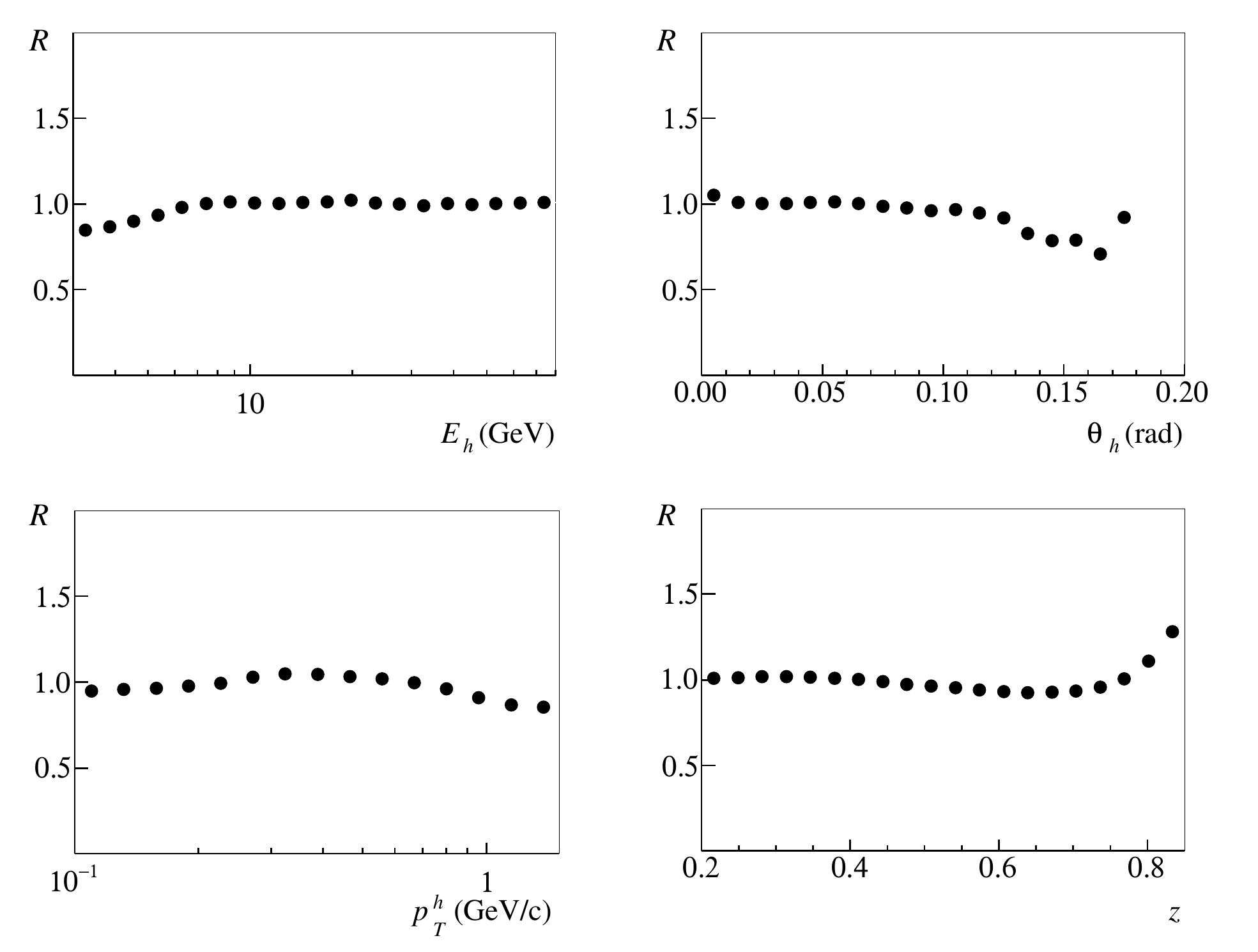} 
\end{center}

\caption{ Ratio $R$ between data and Monte Carlo hadrons distributions
for the energy, the polar angle calculated in the laboratory system, 
$p_T^{\,h}$ and $z$. }
\label{fig:mc_rd_cmp2}
\end{figure}

\begin{figure}[tbh]
\begin{center}
\includegraphics[width=.5\textwidth]{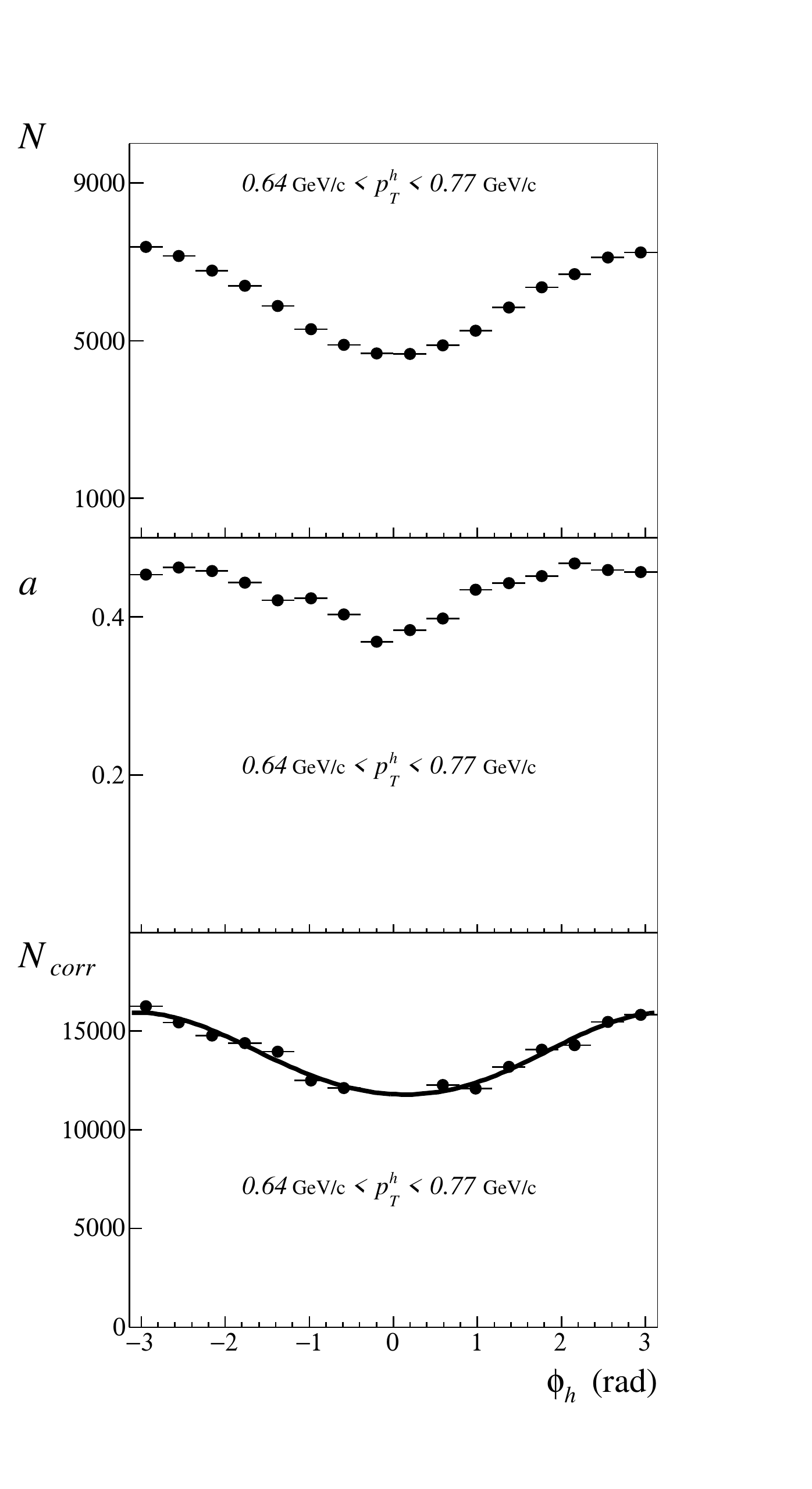}
\end{center}
\caption{\label{fig:hcor} Measured azimuthal distribution $N$, 
azimuthal acceptance $a$ and measured azimuthal distribution corrected 
by the acceptance $N_{corr}$ in one of the $p_T^{\,h}$ bins.}
\end{figure}

The ratios between the distributions for real and for Monte Carlo events are
shown in Fig.~\ref{fig:mc_rd_cmp1} as a function of the DIS variables, and in
Fig.~\ref{fig:mc_rd_cmp2} as a function of the hadron variables. The agreement
is satisfactory and gives confidence in the quality of the apparatus description
used in the simulations. A typical hadron azimuthal distribution from raw data
$N(\phi_h)$, the corresponding acceptance from the Monte Carlo simulation
$a(\phi_h)$, and the corrected distribution $N_{corr}(\phi_h)$ are shown in
Fig.~\ref{fig:hcor} as a function of $\phi_h$.

Equation~(\ref{eq:azi_distr}) shows that the relevant part of the acceptance is
the one containing $\cos\phi_h$, $\cos2\phi_h$ and $\sin\phi_h$ modulations.
The amplitudes of these azimuthal modulations, which are essentially the
corrections given by the Monte Carlo, have been evaluated and their trend has
been studied as a function of the various kinematic variables.  It has been
found that the largest corrections, up to about $15\%$, have to be applied to
the $\cos\phi_h$ modulations. The $\cos2\phi_h$ corrections are of the order of
a few percent and the $\sin\phi_h$ corrections are negligible.

A priory the acceptance function $a(\phi_h,\vec{v})$ evaluated in a particular
bin of a specific variable $x$ could still depend on some geometrical observable
$t$ like the azimuthal or polar angle of the scattered muon or on some other
kinematic variables. It has been verified that this is not the case. When
extracting $a(\phi_h, \vec{v}, t)$ in bins of $t$, the resulting azimuthal
asymmetries differ on average from those extracted through integration over $t$
by less than one standard deviation of the statistical uncertainty, and also
significantly less than the final systematic uncertainty.

\section{Systematic studies}
\label{sec:sys}

Several possible systematic effects have been investigated. The most relevant
studies are described in this section. Some effects turned out to have a
negligible impact on the results and thus were not included in the evaluation of
the final systematic uncertainties.

\subsection{Resolution effects}

Due to the finite resolution of the detectors and of the tracking, the
reconstructed values of some kinematic variables could result in a migration of
an event (or hadron) from one bin to an adjacent bin. This effect can dilute
the measured asymmetries with respect to the true ones. It has been evaluated
using a Monte Carlo event sample with a $\cos\phi_h$ modulation with an
amplitude linearly decreasing as a function of $z$ from $0$ to a value of
$-0.5$. It has been found that the difference between the extracted amplitudes
and the generated ones is always less than $1\%$, and thus it was neglected in
the calculation of the systematic uncertainties.

\subsection{Radiative effects}
\label{sec:rad}

Radiative photons emitted from the lepton modify the reconstructed virtual
photon 4-momentum with respect to the 4-momentum of the true virtual photon
exchanged in the muon-nucleon interaction. This introduces a bias in the
azimuthal distributions, since the reconstructed virtual photon direction in the
lepton scattering plane is always at larger angles than that of the true virtual
photon.

The effect of radiative corrections on the measured asymmetries is expected to
be small for this analysis, because requirement of at least one hadron in the
final state limits the radiative corrections to those for the inelastic part of
the $\gamma^*N$ cross-section. In addition, the use of a muon beam results in
further reduction of radiative corrections. Nevertheless, the effect has been
evaluated by means of Monte Carlo simulations using a dedicated software
(RADGEN~\cite{radgen}) in combination with LEPTO. The correction turns out to
be negligible for the $\cos2\phi_h$ modulation and is small (at most few percent
in the high $x$ region) for the $\cos\phi_h$ modulation, and almost of the same
size for positive and for negative hadrons. The same conclusion has been drawn
by performing an analytic calculation~\cite{rc_cahn_compass} which gives
negligible effects ($\lesssim 1\%$ for the $\cos\phi_h$ modulation) in the
COMPASS environment. For these reasons the radiative corrections have not been
applied to the measured asymmetries and not included in the systematic
uncertainties.

The azimuthal distributions of hadrons are affected by the contamination of
electrons/positrons coming from the conversion of the radiated photons. The
kinematics of the process is such that the contribution is present only in the
two $\phi_h$ bins closest to $\phi_h=0$ ($0 \leq \phi_h < \pi/8$ and $15\pi/8
\leq \phi_h < 2\pi$). In order to avoid corrections depending on the Monte
Carlo description of the radiative effects, these two bins have been excluded in
the extraction of the azimuthal asymmetries.

\subsection{Acceptance corrections}
\label{sec:sys_mc}

\begin{figure}[tbh]
\begin{center}
\includegraphics[width=.8\textwidth]{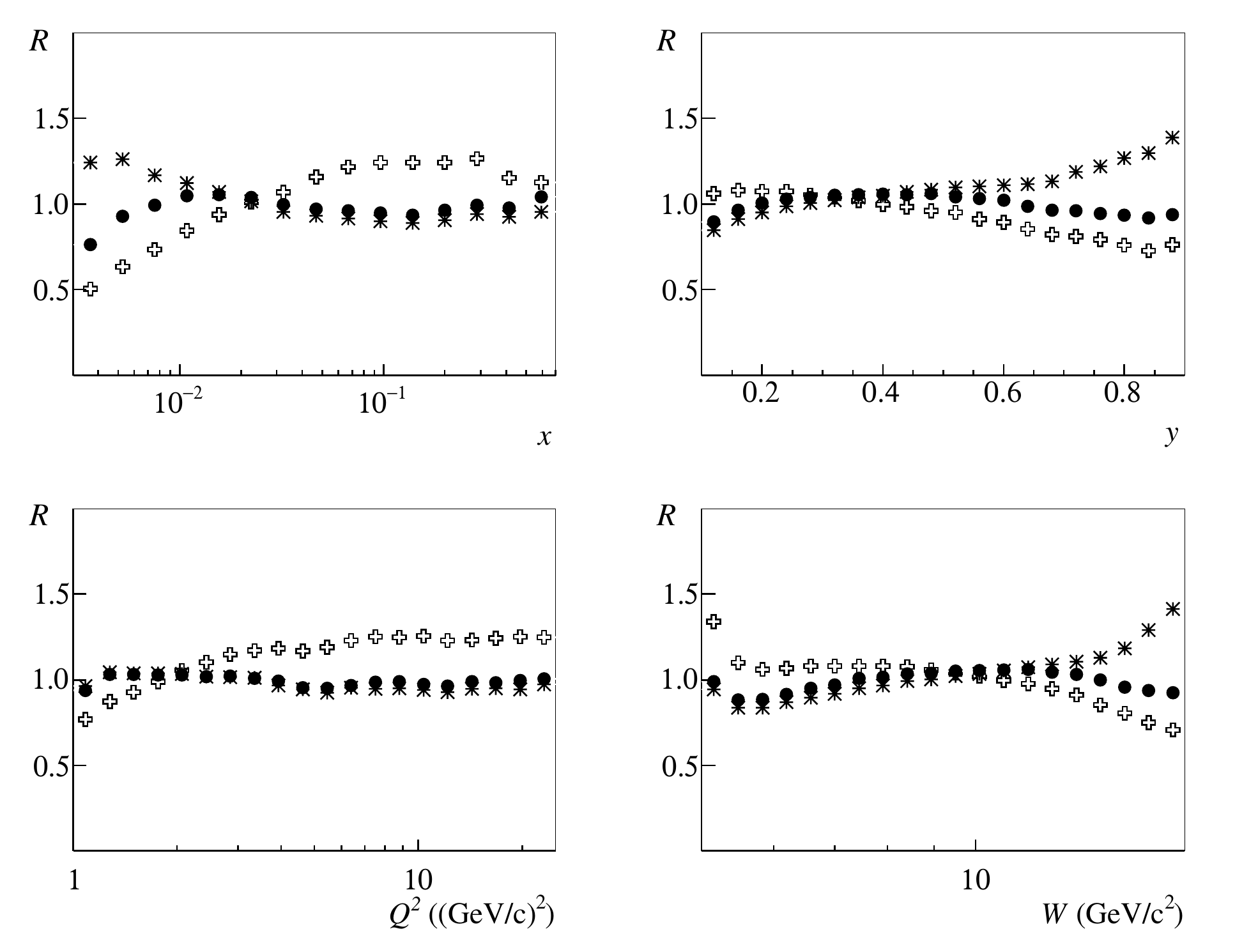} 
\end{center}
\caption{ Ratio $R$ between data and Monte Carlo events distributions. 
The different markers correspond to the three different Monte Carlo 
tunings which have been used to evaluate the acceptance. 
The full points are the same as in Fig.~\ref{fig:mc_rd_cmp1}. }
\label{fig:3mc_cmp}
\end{figure}

\begin{figure}[tbh]
\begin{center}
\includegraphics[width=.99\textwidth]{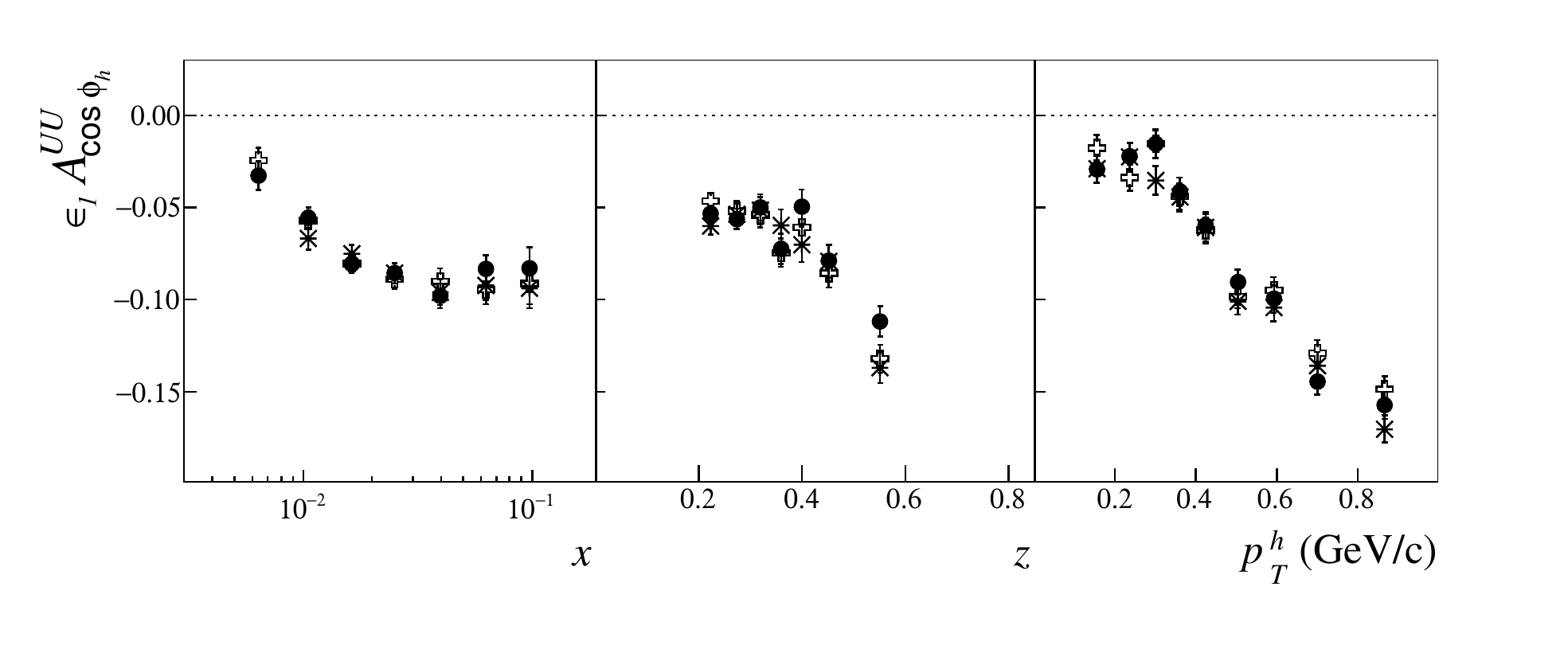} 
\end{center}
\caption{ The amplitudes of the $\cos\phi_h$ modulations 
($\epsilon_1 A^{UU}_{\cos \phi_h}$) 
extracted using the acceptance corrections from the three 
Monte Carlo samples of Fig.~\ref{fig:3mc_cmp}.}
\label{fig:3mc_asy_pos}
\end{figure}

The asymmetries have also been extracted using two other Monte Carlo event
samples. They use the same description of the apparatus but different tuning of
the LEPTO generator. They both compare satisfactorily with the data and can be
considered as ``extreme cases'' as shown in Fig.~\ref{fig:3mc_cmp}. Since the
acceptance is approximately flat in the selected kinematic region the results
are similar as shown for example in Fig.~\ref{fig:3mc_asy_pos}. The difference
between the amplitudes of the azimuthal modulations extracted from the data
corrected with the acceptance calculated using the three different Monte Carlo
samples turned out to be slightly larger than the statistical errors of the
results. These differences have been included in the systematic uncertainties.

\subsection{Stability of the results}
\label{sec:sys_longmc}

The same azimuthal asymmetries have also been extracted from a different data
sample, namely four different weeks of the 2004 run when the target was
longitudinally polarised. A dedicated Monte Carlo simulation has been performed
to describe the apparatus, which was somewhat different from the one used for
the present analysis. The magnetic field in the target region was different and
the beam line was shifted to account for it. Also the triggers were changed and
some detectors parameters were differently tuned. The asymmetries extracted
from these data have been compared with the final ones and the difference
between them (on average, one statistical standard deviation) has been included
in the systematic uncertainties.

\subsection{Detector efficiency}
\label{sec:sys_straw}
A contribution to the azimuthal modulations of the acceptances could be due to
detector inefficiencies in regions where there are less redundancies in the
track reconstruction. A Monte Carlo study has been performed in order to study
the azimuthal modulations of acceptance assuming certain detectors to be
inefficient. The ratio between the azimuthal distributions of the hadrons
reconstructed with reduced efficiency and with nominal detector conditions has
been obtained for every kinematic bin. As a result, it has been found that only
the $\cos\phi_h$ azimuthal modulation changes, in particular in the high $x$
region, where the effect is about $0.03$. This contribution is included in the
systematic uncertainties.

\subsection{Evaluation of the systematic uncertainties}
\label{sec:sys_errors}

The three important contributions to the systematic uncertainties (acceptance
corrections, period compatibility and, to a lesser extent, detector
inefficiencies) have been added up in quadrature and the final systematic
uncertainty $\sigma_{syst}$ has been evaluated to be twice as large as the
statistical ones $\sigma_{stat}$ independently from the kinematic region. This
result, which was obtained in the case of the integrated asymmetries, holds true
also for the 3d asymmetries evaluated in bins of $x$, $z$ and $p_T^{h}$. In
particular, the systematic studies described in Sect.~\ref{sec:sys_mc} which
give the main contribution to the final systematic uncertainty, have been
performed also for the 3d asymmetries.

\section{Results}
\label{sec:results}

\subsection{Asymmetries for separate binning in $x$, $z$ or $p_T^{h}$}
\begin{figure}[tbh]
\begin{center}
\includegraphics[width=.9\textwidth]{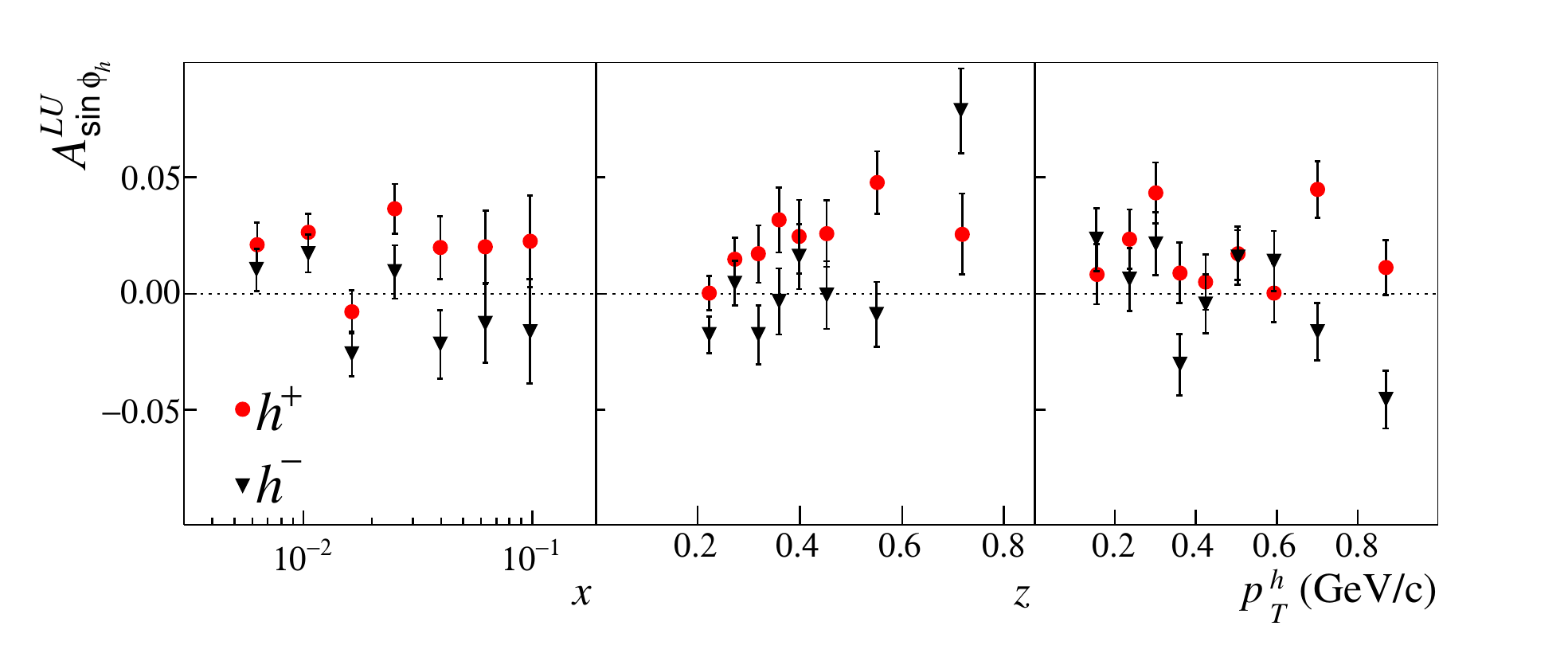} 
\end{center}
\caption{$A^{LU}_{\sin \phi_h}$ integrated asymmetries for positive (red points) 
and negative (black triangles) hadrons as functions of $x$, $z$ 
and $p_T^{\,h}$. The error bars show statistical uncertainties only. }
\label{fig:1d_final_sinphi}
\end{figure}

\begin{figure}[tbh]
\begin{center}
\includegraphics[width=.9\textwidth]{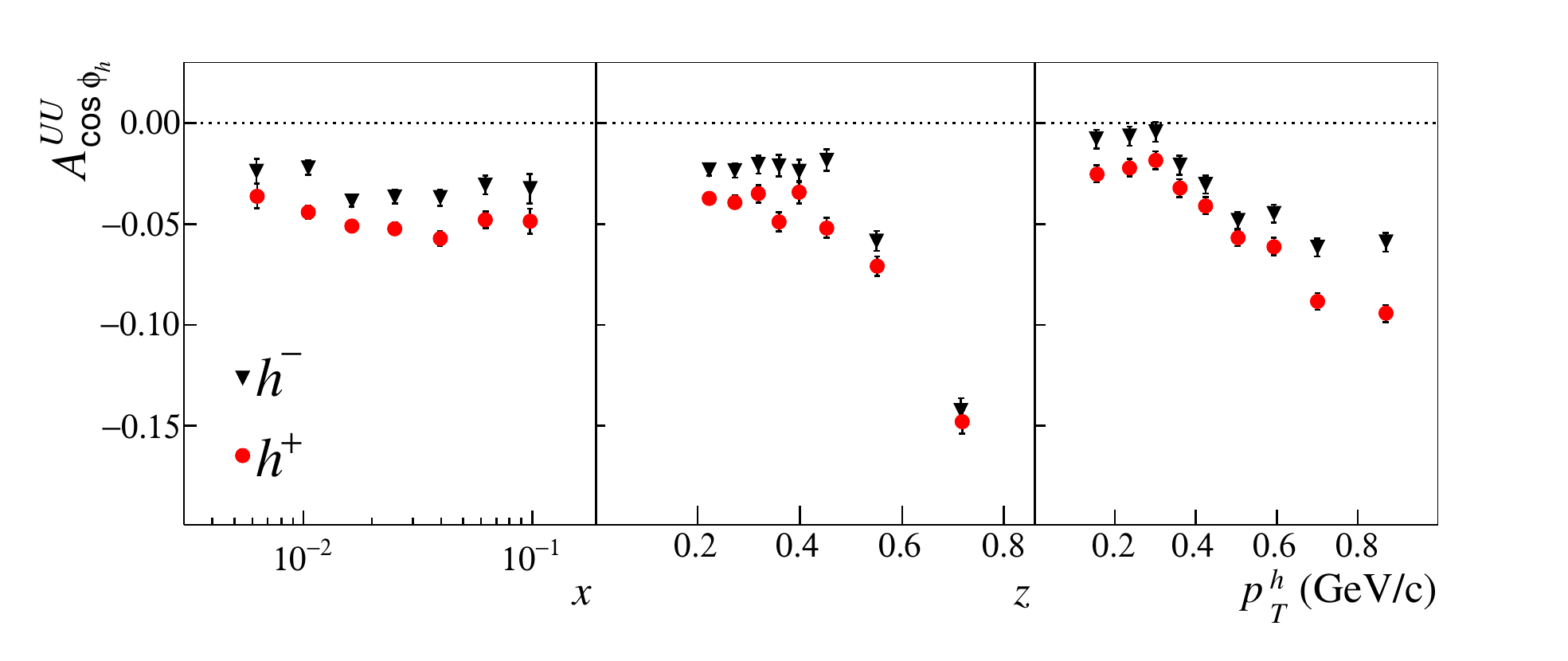} 
\end{center}
\caption{$A^{UU}_{\cos \phi_h}$ integrated asymmetries for positive (red points) 
and for negative (black triangles) hadrons as functions of $x$, $z$ 
and $p_T^{\,h}$. The error bars show statistical uncertainties only. }
\label{fig:1d_final_cosphi}
\end{figure}

\begin{figure}[tbh]
\begin{center}
\includegraphics[width=.9\textwidth]{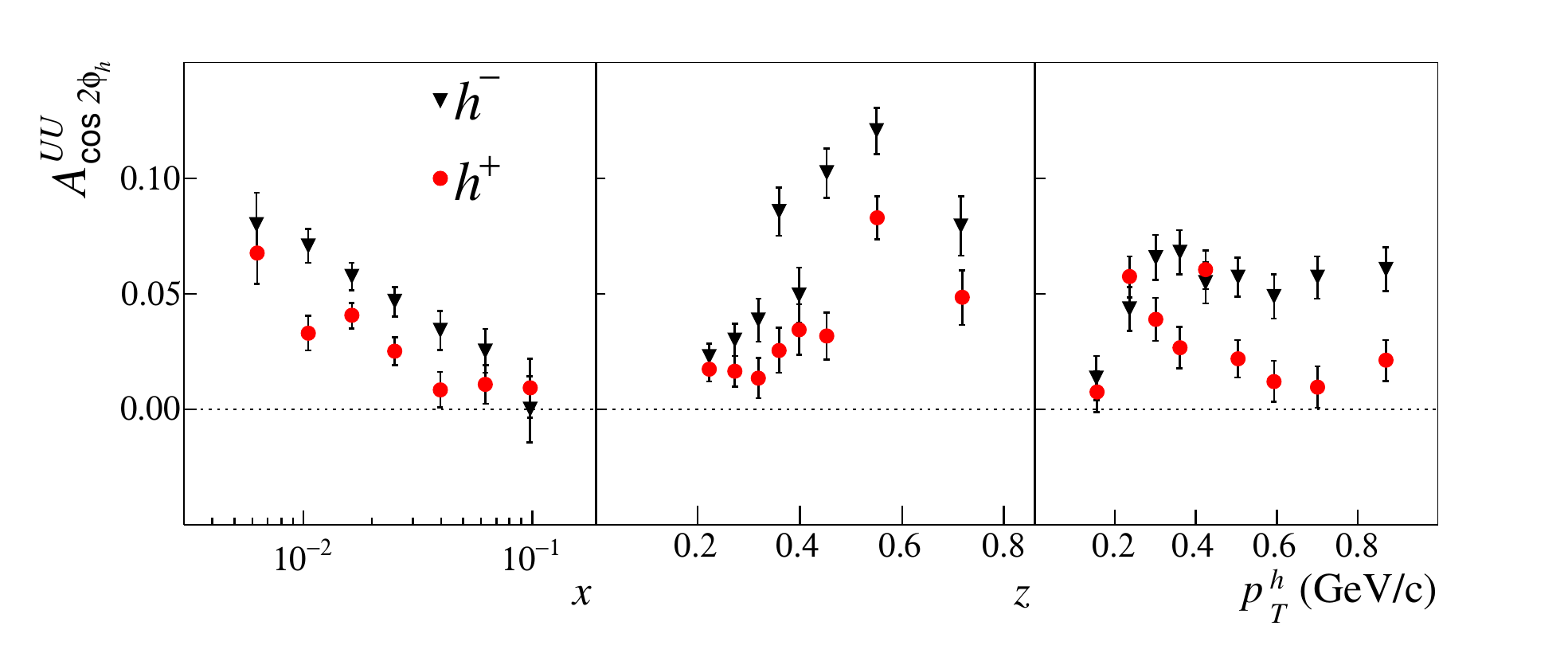} 
\end{center}
\caption{ $A^{UU}_{\cos 2\phi_h}$ integrated asymmetries for positive (red
 points) and negative (black triangles) hadrons as functions of $x$, $z$ and
 $p_T^{\,h}$. The error bars show statistical uncertainties only. }
\label{fig:1d_final_cos2phi}
\end{figure}

The results obtained binning the data in the kinematic variables $x$, $z$ or
$p_T^{h}$ (integrated asymmetries) are listed in Tables 2--4 and shown in
Fig.~\ref{fig:1d_final_sinphi} for $A^{LU}_{\sin \phi_h}$, in
Fig.~\ref{fig:1d_final_cosphi} for $A^{UU}_{\cos \phi_h}$ and in
Fig.~\ref{fig:1d_final_cos2phi} for $A^{UU}_{\cos 2\phi_h}$. The red points and
the black triangles show the asymmetries for positive and negative hadrons,
respectively. The error bars represent statistical uncertainties. As described
in the previous section, the systematic point-to-point uncertainties are
estimated to be as large as twice the statistical ones.
 
As can be seen in Fig.~\ref{fig:1d_final_sinphi}, the $A^{LU}_{\sin \phi_h}$
asymmetry is small, compatible with zero for negative hadrons. For the positive
ones, the asymmetry is slightly positive, increasing with $z$, and almost
constant in $x$ and $p_T^{h}$ within statistical errors. Similar results were
obtained for $\pi^{+}$ by the CLAS Collaboration~\cite{clas_sin} using an
electron beam of $4.3$~GeV and a proton target, and for charged pions by the
HERMES Collaboration~\cite{hermes_sin} with a $27.6$~GeV positron beam and a
proton target. Given the different targets and the different kinematic regions a
quantitative comparison with the present results is not straightforward.

The $A^{UU}_{\cos \phi_h}$ asymmetry given in Fig.~\ref{fig:1d_final_cosphi} is
large and negative for both positive and negative hadrons, with larger absolute
values for positive hadrons. The dependence on the kinematic variables is
strong, in particular on $z$ and $p_T^{h}$. The asymmetries as a function of
$z$ are almost constant up to $z \simeq 0.5$ and increase in absolute value at
larger $z$ up to 0.15. They show a similar behaviour as a function of
$p_T^{h}$: the asymmetries are almost constant up to $p_T^{h} \simeq 0.4$ and
then increase rapidly in absolute value. The comparison with most of the
existing data is difficult because of the different kinematic ranges. Moreover,
the asymmetries have been measured as functions of different variables and
without charge separation. This is not the case for the recently published
results by the HERMES experiment~\cite{hermes_unpol}, which give the asymmetries
as a function of $x$, $y$, $z$ and $p_T^{h}$ both for proton and deuteron
targets and for charged and identified hadrons. However, a quantitative
comparison is still difficult because HERMES measurements correspond to smaller
$Q^2$ values and to larger $x$, although the $x$-dependence of $A^{UU}_{\cos
 \phi_h}$ asymmetry from HERMES is in qualitative agreement with the present
measurement. In particular, the HERMES results also show larger (and negative)
asymmetries for positive hadrons in the overlapping $x$ region. Because of the
different $x$ range, the HERMES mean values are smaller but the data show $z$
and $p_T^{h}$ dependencies similar to that shown in
Fig.~\ref{fig:1d_final_cosphi}. When compared to theoretical calculations and
predictions~\cite{boglione_bis}, the agreement is not satisfactory, in
particular for the $z$ and $p_T^{h}$ dependencies and work to understand the
discrepancies is ongoing.

The $A^{UU}_{\cos 2\phi_h}$ asymmetries are also significantly different from
zero and different for positive and negative hadrons. They both are positive
and larger for negative hadrons, over all the measured range. Again there is a
strong dependence on the kinematic variables. In this case the asymmetry
decreases with $x$ and it increases as functions of $z$ and $p_T^{h}$, but only
up to $z \simeq 0.6$ and $p_T^{h} \simeq 0.4$. Strong dependencies on the
kinematic variables are also present in the HERMES results~\cite{hermes_unpol}.
First attempts to describe the observed behaviour in terms of the Cahn effect
that is expected to dominate at small $x$ and the B-M effect~\cite{barone} could
not reproduce the data well, and in particular the $p_T^{h}$ dependence (the
preliminary results were even not included in the fit) which was expected to be
almost linear.

\subsection{Asymmetries for simultaneous
binning in $x$, $z$ and $p_T^{h}$}

\begin{figure}[tbh]
\begin{center}
\includegraphics[width=0.99\textwidth]{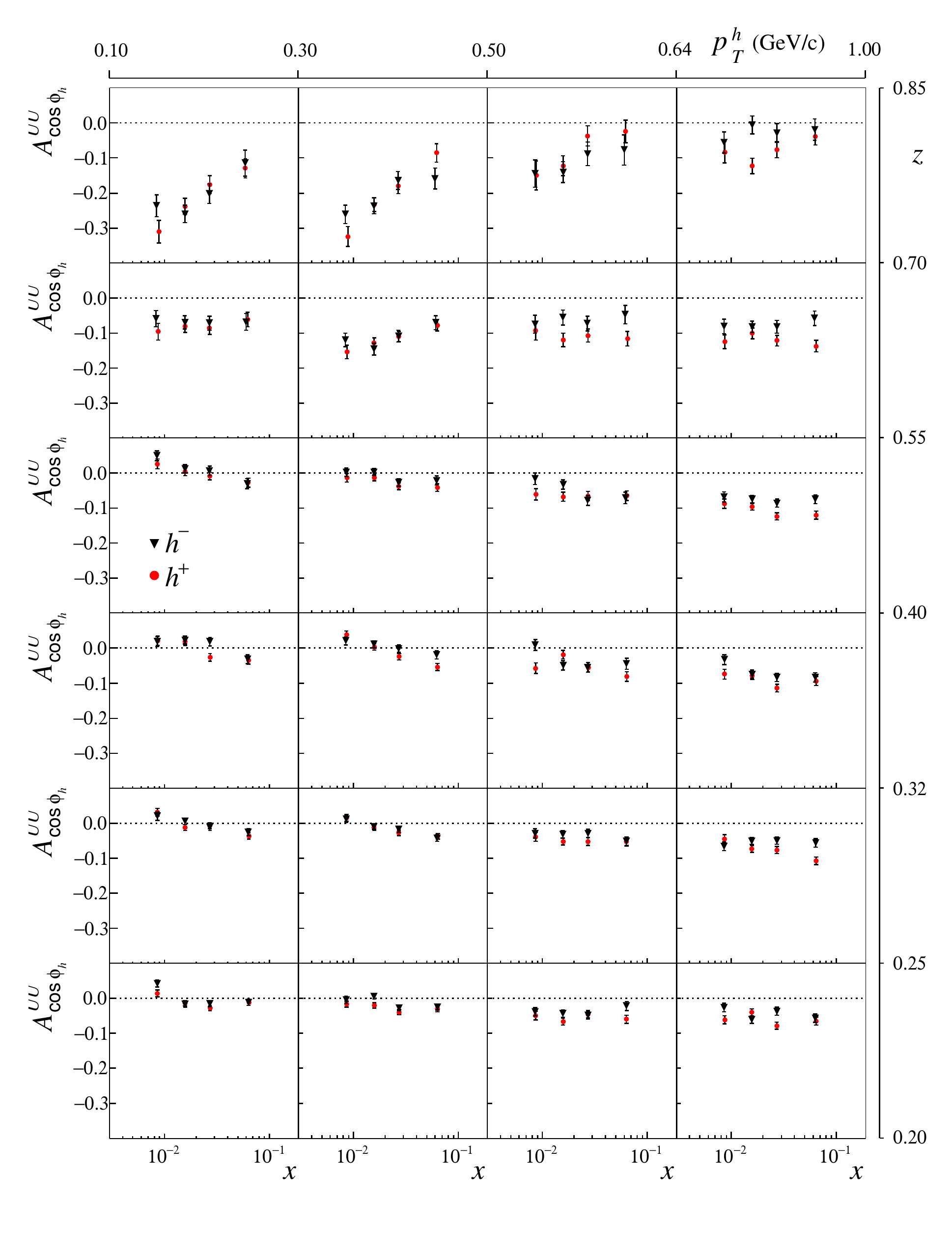} 
\end{center}
\caption{$A^{UU}_{\cos \phi_h}$ asymmetries for positive (red points) and
 negative (black triangles) hadrons as a function of $x$ for the different bins
 in $p_T^{\,h}$ (from left to right) and $z$ (from bottom to top). The error
 bars show statistical uncertainties only. }
\label{fig:3d_Xcosphi}
\end{figure}

\begin{figure}[tbh]
\begin{center}
\includegraphics[width=.99\textwidth]{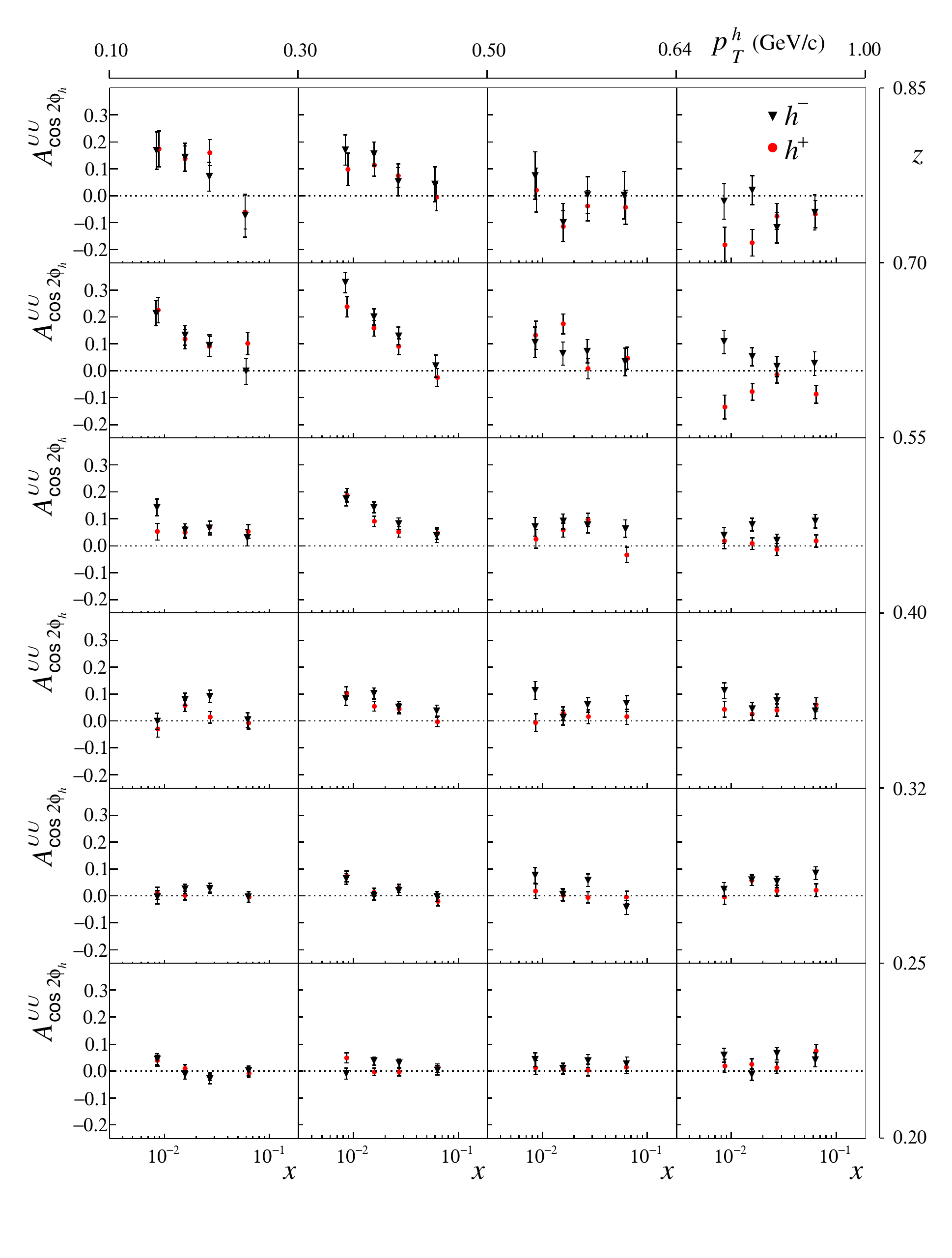} 
\end{center}
\caption{ $A^{UU}_{\cos 2\phi_h}$ asymmetries for positive (red points) and
 negative (black triangles) hadrons as a function of $x$ for the different bins
 in $p_T^{\,h}$ (from left to right) and $z$ (from bottom to top).The error
 bars show statistical uncertainties only. }
\label{fig:3d_Xcos2phi}
\end{figure}

In order to investigate the observed dependencies on kinematic variables, the
azimuthal asymmetries have also been extracted binning simultaneously the data
in bins of $x$, $z$ and $p_T^{\,h}$ (3d asymmetries). The results for the four
$x$ bins are given in Tables 5--12. The results for $A^{UU}_{\cos \phi_h}$ for
positive (red points) and negative (black triangles) hadrons are shown in
Fig.~\ref{fig:3d_Xcosphi}. The results for $A^{UU}_{\cos 2\phi_h}$ are shown in
Fig.~\ref{fig:3d_Xcos2phi} and again the error bars represent only the
statistical uncertainties. The 3d asymmetries have also been evaluated for
$A^{LU}_{\sin \phi_h}$ but no particular effect could be noticed due to the
larger statistical uncertainties. It has also been checked that the projection
of the asymmetries on any of the three kinematic variables is consistent with
the results for the integrated asymmetries given in the previous section.

From the results shown in Fig.~\ref{fig:3d_Xcosphi} an interesting information
on $A^{UU}_{\cos \phi_h}$ can be obtained. Looking at the $x$ dependence in the
$z$ and $p_T^{\,h}$ bins, it is clear that the large negative values at small
$x$ are mostly due to the hadrons with $0.55 < z < 0.85$, while for smaller $z$
the asymmetries are either very small ($0.1$~GeV/$c < p_T^{\,h} < 0.5$~GeV/$c$)
or indicate a different $x$ dependence ($p_T^{\,h} > 0.5$~GeV/$c$). Also, as can
be seen in the figure, the absolute values of the asymmetries for $z < 0.55$
increase somewhat with $p_T^{\,h}$, and the large and negative values at large
$z$ are mainly due to the values at small $x$ and $p_T^{\,h}$. Summarising, the
data suggest that there are different regimes and different dominant processes
in the various regions of the ($z$, $p_T^{\,h}$) plane and that a deeper
phenomenological investigation is required.

Also $A^{UU}_{\cos 2\phi_h}$ shows a similarly strong dependence on the $x$, $z$
and $p_T^{\,h}$ variables, as can be seen in Fig.~\ref{fig:3d_Xcos2phi}. The
large positive asymmetry values in the low-$x$ region are mainly observed at
small $p_T^{\,h}$ values and large $z$ values. For $p_T^{\,h} > 0.5$~GeV/$c$
$A^{UU}_{\cos 2\phi_h}$ becomes smaller and shows a different $x$ dependence.

\begin{figure}[tbh]
\begin{center}
\includegraphics[width=.8\textwidth]{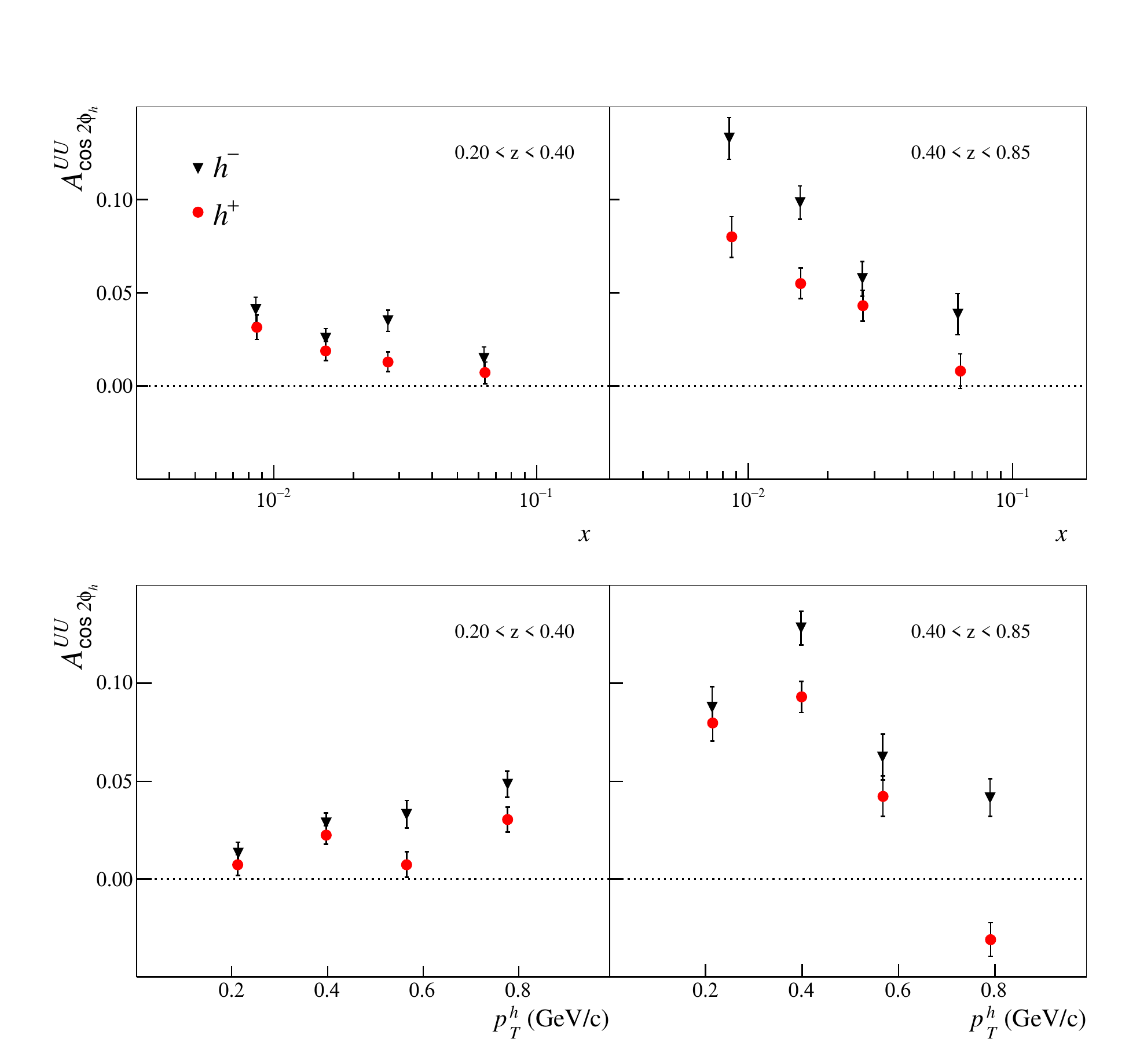} 
\end{center}
\caption{ $A^{UU}_{\cos \phi_h}$ as functions of $x$ (top) and $p_T^{\,h}$
 (bottom) calculated for $0.2 < z <0.4$ (left) and for $0.4 < z < 0.85$
 (right).The error bars show statistical uncertainties only. }
\label{fig:zrange_cosphi}
\end{figure}

\begin{figure}[tbh]
\begin{center}
\includegraphics[width=.8\textwidth]{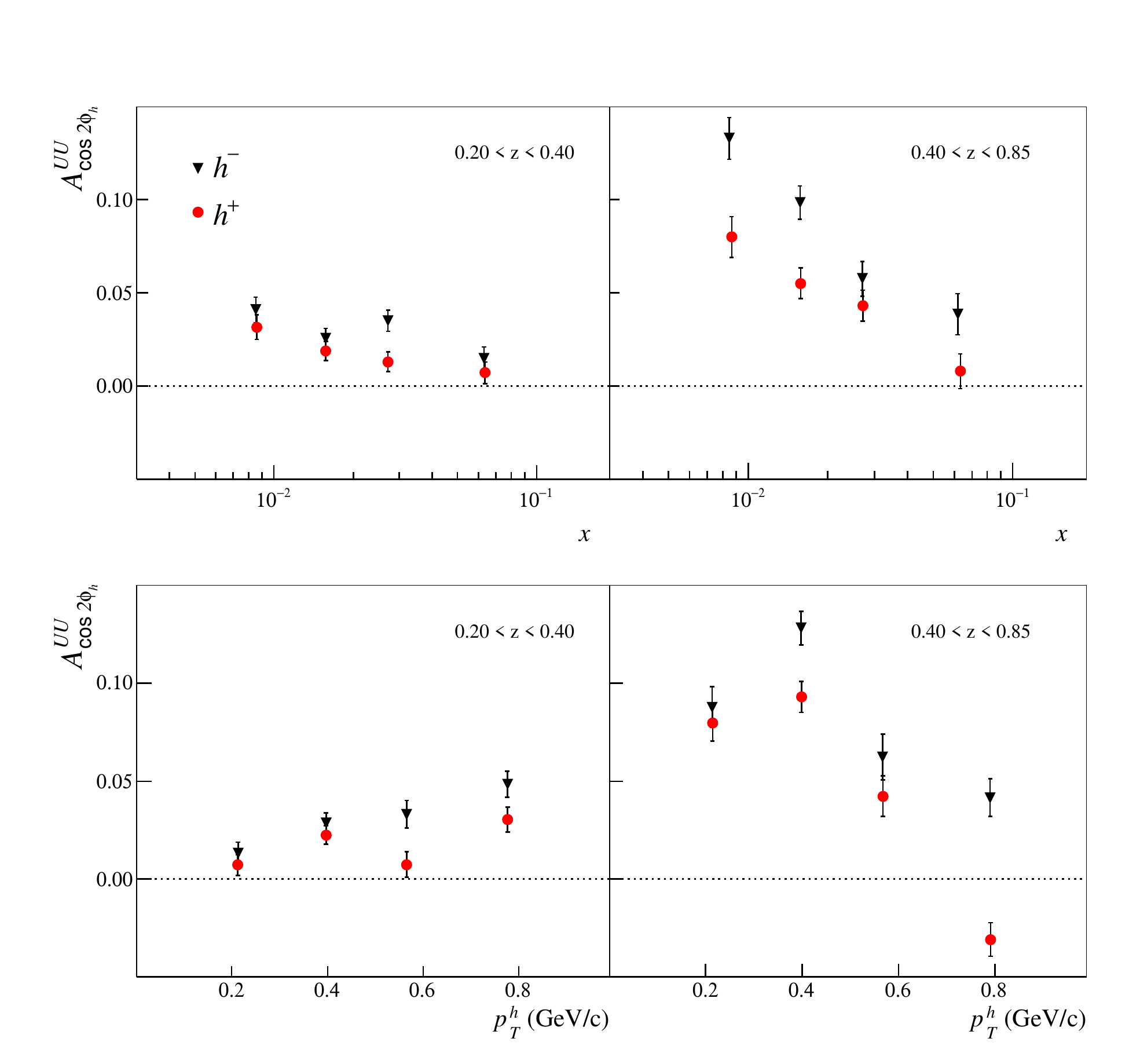} 
\end{center}
\caption{ $A^{UU}_{\cos 2\phi_h}$ as functions of $x$ (top) and $p_T^{\,h}$
 (bottom) calculated for $0.2 < z <0.4$ (left) and for $0.4 < z < 0.85$
 (right).The error bars show statistical uncertainties only. }
\label{fig:zrange_cos2phi}
\end{figure}

The presence of two different regimes according to the $z$ values appears very
clearly in Fig.~\ref{fig:zrange_cosphi} and Fig.~\ref{fig:zrange_cos2phi}. Here
the asymmetries have been calculated at low $z$ ($0.2 < z < 0.4$) and at high
$z$ ($0.4 < z < 0.85$) and both the $x$ and $p_T^{\,h}$ dependencies of the
$A^{UU}_{\cos \phi_h}$ and $A^{UU}_{\cos 2\phi_h}$ asymmetries are found to be
significantly different. As a side remark we can remind that the low $z$
behaviour is qualitatively reproduced by the existing fit and
calculations~\cite{barone,boglione_bis}, while the dependence at high $z$ seems
to be more difficult to be reproduced theoretically.

\section{Conclusions}
\label{conclusions}

COMPASS has measured the azimuthal asymmetries in SIDIS of $160$~GeV/$c$ muons
off an unpolarised isoscalar target, covering a broad $x$ region down to $x=0.003$.
Results have been produced binning the data in $x$, $z$ or $p_T^{h}$,
integrating over the other two variables, and in a three-dimensional grid of the
three variables $x$, $z$ and $p_T^{h}$. The dependencies of the amplitudes of
the $\cos\phi_h$ and $\cos2\phi_h$ modulations over the kinematic variables turn
out to be very strong and not easy to be described in the present
phenomenological framework. The new data can be used in multidimensional global
analyses and constitute an important information for the understanding of the
transverse momentum structure of the nucleon.

\textbf{Acknowledgements}
We gratefully acknowledge the support of the CERN management and staff and the
skill and effort of the technicians of our collaborating institutes. This work
was made possible by the financial support of our funding agencies. We would
like to thank A. Afanasev, V. Barone, M. Boglione, and S. Melis, for the many
fruitful discussions.
\FloatBarrier
\clearpage

\begin{table*}
\caption{$A^{UU}_{\sin \phi_h}$ asymmetries for positive and negative hadrons as
 functions of $x$, $z$ and $p_T^{\,h}$.
The errors are statistical only.}
\label{tab:values_1d_s}
\begin{center}
\begin{tabular}{  c  r  r  }
\hline
$x$ range & $A^{UU}_{\sin \phi_h}$ (h$^+$) & $A^{UU}_{\sin \phi_h}$ (h$^-$) \\
\hline
$0.003 - 0.008$ & $ 0.021 \pm 0.009$ & $  0.010 \pm  0.009$ \\
$0.008 - 0.013$ & $ 0.026 \pm 0.008$ & $  0.017 \pm  0.008$ \\
$0.013 - 0.020$ & $ -0.007 \pm 0.009$ & $ -0.026 \pm  0.010$ \\
$0.020 - 0.032$ & $ 0.036 \pm 0.011$ & $  0.009 \pm  0.011$ \\
$0.032 - 0.050$ & $ 0.020 \pm 0.013$ & $ -0.022 \pm  0.015$ \\
$0.050 - 0.080$ & $ 0.020 \pm 0.015$ & $ -0.013 \pm  0.017$ \\
$0.080 - 0.130$ & $ 0.022 \pm 0.019$ & $ -0.016 \pm  0.022$ \\
\hline
$z$ range & $A^{UU}_{\sin \phi_h}$ (h$^+$) & $A^{UU}_{\sin \phi_h}$ (h$^-$) \\
\hline
$0.20 - 0.25$  & $ 0.000 \pm 0.007$ & $ -0.018 \pm 0.008 $ \\
$0.25 - 0.30$  & $ 0.015 \pm 0.009$ & $  0.004 \pm 0.009 $ \\
$0.30 - 0.34$  & $ 0.017 \pm 0.012$ & $ -0.018 \pm 0.013 $ \\
$0.34 - 0.38$  & $ 0.032 \pm 0.014$ & $ -0.003 \pm 0.014 $ \\
$0.38 - 0.42$  & $ 0.024 \pm 0.016$ & $  0.016 \pm 0.014 $ \\
$0.42 - 0.49$  & $ 0.026 \pm 0.014$ & $ -0.001 \pm 0.015 $ \\
$0.49 - 0.63$  & $ 0.048 \pm 0.013$ & $ -0.009 \pm 0.014 $ \\
$0.63 - 0.85$  & $ 0.025 \pm 0.017$ & $  0.078 \pm 0.018 $ \\
\hline
$p_T^{\,h}$ range {\small (GeV/$c$)} & $A^{UU}_{\sin \phi_h}$ (h$^+$)& $A^{UU}_{\sin \phi_h}$ (h$^-$) \\
\hline
$0.10 - 0.20 $ & $ 0.008 \pm 0.013 $ & $  0.023 \pm 0.013 $ \\
$0.20 - 0.27 $ & $ 0.023 \pm 0.013 $ & $  0.006 \pm 0.013 $ \\
$0.27 - 0.33 $ & $ 0.043 \pm 0.013 $ & $  0.021 \pm 0.013 $ \\
$0.33 - 0.39 $ & $ 0.009 \pm 0.013 $ & $ -0.030 \pm 0.013 $ \\
$0.39 - 0.46 $ & $ 0.005 \pm 0.012 $ & $ -0.004 \pm 0.013 $ \\
$0.46 - 0.55 $ & $ 0.017 \pm 0.011 $ & $  0.015 \pm 0.012 $ \\
$0.55 - 0.64 $ & $ 0.001 \pm 0.012 $ & $  0.014 \pm 0.013 $ \\
$0.64 - 0.77 $ & $ 0.045 \pm 0.012 $ & $ -0.016 \pm 0.012 $ \\
$0.77 - 1.00 $ & $ 0.011 \pm 0.012 $ & $ -0.045 \pm 0.012 $ \\
\hline
\end{tabular}
\end{center}
\end{table*}

\begin{table*}
\caption{$A^{UU}_{\cos \phi_h}$ asymmetries for positive and negative hadrons as
 functions of $x$, $z$ and $p_T^{\,h}$.
The errors are statistical only.}
\label{tab:values_1d_c}
\begin{center}
\begin{tabular}{  c  r  r  }
\hline
$x$ range & $A^{UU}_{\cos \phi_h}$ (h$^+$) & $A^{UU}_{\cos \phi_h}$ (h$^-$) \\
\hline
$0.003 - 0.008 $ & $-0.036  \pm 0.006 $ & $ -0.024 \pm 0.006$\\
$0.008 - 0.013 $ & $-0.044  \pm 0.003 $ & $ -0.022 \pm 0.003$\\
$0.013 - 0.020 $ & $-0.051  \pm 0.003 $ & $ -0.039 \pm 0.003$\\
$0.020 - 0.032 $ & $-0.052  \pm 0.003 $ & $ -0.037 \pm 0.003$\\
$0.032 - 0.050 $ & $-0.057  \pm 0.004 $ & $ -0.037 \pm 0.004$\\
$0.050 - 0.080 $ & $-0.048  \pm 0.004 $ & $ -0.031 \pm 0.005$\\
$0.080 - 0.130 $ & $-0.048  \pm 0.006 $ & $ -0.032 \pm 0.007$\\
\hline
$z$ range & $A^{UU}_{\cos \phi_h}$ (h$^+$) & $A^{UU}_{\cos \phi_h}$ (h$^-$) \\
\hline
$0.20 - 0.25 $ & $ -0.037 \pm 0.003 $ & $  -0.023 \pm 0.003$\\
$0.25 - 0.30 $ & $ -0.039 \pm 0.003 $ & $  -0.024 \pm 0.003$\\
$0.30 - 0.34 $ & $ -0.035 \pm 0.004 $ & $  -0.020 \pm 0.004$\\
$0.34 - 0.38 $ & $ -0.049 \pm 0.005 $ & $  -0.021 \pm 0.005$\\
$0.38 - 0.42 $ & $ -0.034 \pm 0.005 $ & $  -0.024 \pm 0.006$\\
$0.42 - 0.49 $ & $ -0.052 \pm 0.005 $ & $  -0.018 \pm 0.005$\\
$0.49 - 0.63 $ & $ -0.071 \pm 0.005 $ & $  -0.058 \pm 0.005$\\
$0.63 - 0.85 $ & $ -0.148 \pm 0.006 $ & $  -0.142 \pm 0.006$\\
\hline
$p_T^{\,h}$ range {\small (GeV/$c$)} & $A^{UU}_{\cos \phi_h}$ (h$^+$) & $A^{UU}_{\cos \phi_h}$ (h$^-$)\\
\hline
$0.10 - 0.20 $ &$  -0.025 \pm  0.004$  & $ -0.008 \pm 0.005$\\
$0.20 - 0.27 $ &$  -0.022 \pm  0.004$  & $ -0.006 \pm 0.005$\\
$0.27 - 0.33 $ &$  -0.018 \pm  0.004$ & $ -0.004 \pm 0.005$\\
$0.33 - 0.39 $ &$  -0.032 \pm  0.004$ & $ -0.020 \pm 0.005$\\
$0.39 - 0.46 $ &$  -0.041 \pm  0.004$ & $ -0.030 \pm 0.004$\\
$0.46 - 0.55 $ &$  -0.057 \pm  0.004$ & $ -0.048 \pm 0.004$\\
$0.55 - 0.64 $ &$  -0.061 \pm  0.004$ & $ -0.045 \pm 0.005$\\
$0.64 - 0.77 $ &$  -0.088 \pm  0.004$ & $ -0.062 \pm 0.004$\\
$0.77 - 1.00 $ &$  -0.094 \pm  0.004$ & $ -0.059 \pm 0.005$\\
\hline
\end{tabular}
\end{center}
\end{table*}

\begin{table*}[htb]
\caption{$A^{UU}_{\cos 2 \phi_h}$ asymmetries for positive and negative hadrons
 as functions of $x$, $z$ and $p_T^{\,h}$. The errors are statistical only.}
\label{tab:values_1d_c2}
\begin{center}
\begin{tabular}{  c  r  r  }
\hline
$x$ range & $A^{UU}_{\cos2 \phi_h}$ (h$^+$) & $A^{UU}_{\cos2 \phi_h}$ (h$^-$) \\
\hline
$0.003 - 0.008 $ & $ 0.068 \pm 0.014 $& $  0.080 \pm 0.014 $ \\
$0.008 - 0.013 $ & $ 0.033 \pm 0.007 $& $  0.071 \pm 0.007 $\\
$0.013 - 0.020 $ & $ 0.040 \pm 0.006 $& $  0.057 \pm 0.006 $\\
$0.020 - 0.032 $ & $ 0.025 \pm 0.006 $& $  0.046 \pm 0.006 $\\
$0.032 - 0.050 $ & $ 0.008 \pm 0.007 $& $  0.034 \pm 0.008 $\\
$0.050 - 0.080 $ & $ 0.011 \pm 0.008 $& $  0.025 \pm 0.009 $\\
$0.080 - 0.130 $ & $ 0.009 \pm 0.013 $& $  0.000 \pm 0.014 $\\
\hline
$z$ range & $A^{UU}_{\cos2 \phi_h}$ (h$^+$) & $A^{UU}_{\cos2 \phi_h}$ (h$^-$) \\
\hline
$0.20 - 0.25 $ & $ 0.017 \pm 0.005 $ & $  0.023 \pm 0.006 $\\
$0.25 - 0.30 $ & $ 0.016 \pm 0.006 $ & $  0.030 \pm 0.007 $\\
$0.30 - 0.34 $ & $ 0.013 \pm 0.009 $ & $  0.038 \pm 0.009 $\\
$0.34 - 0.38 $ & $ 0.025 \pm 0.010 $ & $  0.085 \pm 0.010 $\\
$0.38 - 0.42 $ & $ 0.034 \pm 0.011 $ & $  0.049 \pm 0.012 $\\
$0.42 - 0.49 $ & $ 0.032 \pm 0.010 $ & $  0.102 \pm 0.011 $\\
$0.49 - 0.63 $ & $ 0.083 \pm 0.009 $ & $  0.120 \pm 0.010 $\\
$0.63 - 0.85 $ & $ 0.048 \pm 0.012 $ & $  0.079 \pm 0.013 $\\
\hline
$p_T^{\,h}$ range {\small (GeV/$c$)} & $A^{UU}_{\cos2 \phi_h}$ (h$^+$) & $A^{UU}_{\cos2 \phi_h}$ (h$^-$) \\
\hline
$0.10 - 0.20 $ & $ 0.008 \pm 0.009 $ & $  0.014 \pm 0.009 $ \\
$0.20 - 0.27 $ & $ 0.057 \pm 0.009 $ & $  0.043 \pm 0.010 $ \\
$0.27 - 0.33 $ & $ 0.039 \pm 0.009 $ & $  0.065 \pm 0.010 $ \\
$0.33 - 0.39 $ & $ 0.027 \pm 0.009 $ & $  0.068 \pm 0.010 $ \\
$0.39 - 0.46 $ & $ 0.060 \pm 0.008 $ & $  0.055 \pm 0.009 $ \\
$0.46 - 0.55 $ & $ 0.022 \pm 0.008 $ & $  0.057 \pm 0.008 $ \\
$0.55 - 0.64 $ & $ 0.012 \pm 0.009 $ & $  0.049 \pm 0.009 $ \\
$0.64 - 0.77 $ & $ 0.009 \pm 0.009 $ & $  0.057 \pm 0.009 $ \\
$0.77 - 1.00 $ & $ 0.021 \pm 0.009 $ & $  0.060 \pm 0.009 $ \\
\hline
\end{tabular}
\end{center}
\end{table*}

\begin{table*}[htb]
\caption{$A^{UU}_{\cos \phi_h}$ 3d asymmetries asymmetries for positive and negative hadrons in the first $x$
bin (0.003$< x <$0.012).
The errors are statistical only.}
\label{tab:values_3d_c_1}
\begin{center}
\begin{tabular}{ c c r r  }
\hline
$z$ range & $p_T^{\,h}$ range {\small (GeV/$c$)}& $A^{UU}_{\cos \phi_h}$(h$^+$) &  $A^{UU}_{\cos \phi_h}$(h$^-$) \\
\hline
$0.20-0.25$ & $0.10 - 0.30$ &$ 0.014 \pm 0.010 $ & $ 0.041 \pm 0.010 $  \\
      & $0.30 - 0.50$ &$ -0.017 \pm 0.009 $ & $-0.005 \pm 0.010 $  \\	
      & $0.50 - 0.64$ &$ -0.050 \pm 0.012 $ & $-0.039 \pm 0.012 $  \\	
      & $0.64 - 1.00$ &$ -0.062 \pm 0.011 $ & $-0.026 \pm 0.012 $  \\	
\hline			   			     		   
$0.25-0.32$ & $0.10 - 0.30$ &$ 0.031 \pm 0.011 $ & $ 0.020 \pm 0.011 $  \\  
      & $0.30 - 0.50$ &$ 0.016 \pm 0.009 $ & $ 0.012 \pm 0.010 $  \\   	
      & $0.50 - 0.64$ &$ -0.039 \pm 0.013 $ & $-0.030 \pm 0.013 $  \\  	
      & $0.64 - 1.00$ &$ -0.045 \pm 0.012 $ & $-0.066 \pm 0.012 $  \\  
\hline			   			     		   
$0.32-0.40$ & $0.10 - 0.30$ &$ 0.021 \pm 0.013 $ & $ 0.019 \pm 0.014 $  \\ 
      & $0.30 - 0.50$ &$ 0.038 \pm 0.012 $ & $ 0.021 \pm 0.012 $  \\ 	
      & $0.50 - 0.64$ &$ -0.058 \pm 0.015 $ & $ 0.009 \pm 0.016 $  \\ 	
      & $0.64 - 1.00$ &$ -0.074 \pm 0.014 $ & $-0.033 \pm 0.014 $  \\	
\hline			   			     		   
$0.40-0.55$ & $0.10 - 0.30$ &$ 0.026 \pm 0.014 $ & $ 0.050 \pm 0.014 $  \\ 
      & $0.30 - 0.50$ &$ -0.013 \pm 0.012 $ & $ 0.003 \pm 0.012 $  \\ 
      & $0.50 - 0.64$ &$ -0.061 \pm 0.016 $ & $-0.016 \pm 0.017 $  \\
      & $0.64 - 1.00$ &$ -0.087 \pm 0.014 $ & $-0.068 \pm 0.015 $  \\
\hline			   	    	 	     	       
$0.55-0.70$ & $0.10 - 0.30$ &$ -0.096 \pm 0.024 $ & $-0.059 \pm 0.023 $  \\
      & $0.30 - 0.50$ &$ -0.154 \pm 0.019 $ & $-0.120 \pm 0.019 $  \\
      & $0.50 - 0.64$ &$ -0.093 \pm 0.026 $ & $-0.074 \pm 0.026 $  \\
      & $0.64 - 1.00$ &$ -0.124 \pm 0.020 $ & $-0.080 \pm 0.020 $  \\
\hline			     		     		   
$0.70-0.85$ & $0.10 - 0.30$ &$ -0.310 \pm 0.032 $ & $-0.236 \pm 0.031 $  \\
      & $0.30 - 0.50$ &$ -0.325 \pm 0.029 $ & $-0.260 \pm 0.026 $  \\
      & $0.50 - 0.64$ &$ -0.149 \pm 0.041 $ & $-0.145 \pm 0.039 $  \\
      & $0.64 - 1.00$ &$ -0.083 \pm 0.031 $ & $-0.056 \pm 0.031 $  \\

\hline
\end{tabular}
\end{center}
\end{table*}

\begin{table*}[htb]
\caption{$A^{UU}_{\cos \phi_h}$ 3d asymmetries asymmetries for positive and negative hadrons in the second $x$
bin (0.012$<x<$0.020).
The errors are statistical only.}
\label{tab:values_3d_c_2}
\begin{center}
\begin{tabular}{ c c r r  }
\hline
$z$ range  &  $p_T^{\,h}$ range {\small (GeV/$c$)} &   $A^{UU}_{\cos \phi_h}$(h$^+$)  &  $A^{UU}_{\cos \phi_h}$(h$^-$) \\
\hline
$0.20-0.25$ & $0.10 - 0.30$ & $ -0.014 \pm 0.008 $ &$  -0.017 \pm 0.008$\\
      & $0.30 - 0.50$ & $ -0.021 \pm 0.007 $ &$  0.005 \pm 0.007$\\ 
      & $0.50 - 0.64$ & $ -0.067 \pm 0.010 $ &$  -0.044 \pm 0.010$\\
      & $0.64 - 1.00$ & $ -0.039 \pm 0.010 $ &$  -0.061 \pm 0.010$\\
\hline	     	      		     		   
$0.25-0.32$ & $0.10 - 0.30$ & $ -0.012 \pm 0.008 $ &$  0.005 \pm 0.009$\\ 
      & $0.30 - 0.50$ & $ -0.013 \pm 0.007 $ &$  -0.010 \pm 0.008$\\
      & $0.50 - 0.64$ & $ -0.052 \pm 0.010 $ &$  -0.032 \pm 0.011$\\
      & $0.64 - 1.00$ & $ -0.074 \pm 0.010 $ &$  -0.051 \pm 0.010$\\
\hline	     	      		     		   
$0.32-0.40$ & $0.10 - 0.30$ & $ 0.019 \pm 0.011 $ &$  0.023 \pm 0.012$\\ 
      & $0.30 - 0.50$ & $ 0.003 \pm 0.009 $ &$  0.011 \pm 0.010$\\ 
      & $0.50 - 0.64$ & $ -0.019 \pm 0.012 $ &$  -0.049 \pm 0.013$\\
      & $0.64 - 1.00$ & $ -0.078 \pm 0.011 $ &$  -0.075 \pm 0.011$\\
\hline	     	      		     		   
$0.40-0.55$ & $0.10 - 0.30$ & $ 0.004 \pm 0.011 $ &$  0.013 \pm 0.012$\\ 
      & $0.30 - 0.50$ & $ -0.013 \pm 0.010 $ &$  0.003 \pm 0.010$\\ 
      & $0.50 - 0.64$ & $ -0.068 \pm 0.013 $ &$  -0.033 \pm 0.014$\\
      & $0.64 - 1.00$ & $ -0.096 \pm 0.011 $ &$  -0.075 \pm 0.012$\\
\hline	     	      		     		   
$0.55-0.70$ & $0.10 - 0.30$ & $ -0.080 \pm 0.018 $ &$  -0.069 \pm 0.019$\\
      & $0.30 - 0.50$ & $ -0.129 \pm 0.015 $ &$  -0.146 \pm 0.016$\\
      & $0.50 - 0.64$ & $ -0.120 \pm 0.019 $ &$  -0.055 \pm 0.021$\\
      & $0.64 - 1.00$ & $ -0.100 \pm 0.015 $ &$  -0.082 \pm 0.017$\\
\hline	     	      		     		   
$0.70-0.85$ & $0.10 - 0.30$ & $ -0.238 \pm 0.023 $ &$  -0.260 \pm 0.025$\\
      & $0.30 - 0.50$ & $ -0.233 \pm 0.020 $ &$  -0.237 \pm 0.023$\\
      & $0.50 - 0.64$ & $ -0.122 \pm 0.029 $ &$  -0.140 \pm 0.030$\\
      & $0.64 - 1.00$ & $ -0.122 \pm 0.022 $ &$  -0.006 \pm 0.026$\\

\hline
\end{tabular}
\end{center}
\end{table*}

\begin{table*}
\caption{ $A^{UU}_{\cos \phi_h}$ 3d asymmetries asymmetries for positive and negative hadrons in the third $x$
bin (0.020$<x<$0.038). The errors are statistical only.}
\label{tab:values_3d_c_3}
\begin{center}
\begin{tabular}{ c c r r  }
\hline
$z$ range  & $p_T^{\,h}$ range {\small (GeV/$c$)} & $A^{UU}_{\cos \phi_h}$(h$^+$) & $A^{UU}_{\cos \phi_h}$(h$^-$)\\
\hline
$0.20-0.25$ & $ 0.10 - 0.30 $&$ -0.028 \pm 0.008 $ & $ -0.015 \pm 0.009$\\
      & $ 0.30 - 0.50 $&$ -0.039 \pm 0.007 $ & $ -0.028 \pm 0.008$\\
      & $ 0.50 - 0.64 $&$ -0.046 \pm 0.010 $ & $ -0.048 \pm 0.011$\\
      & $ 0.64 - 1.00 $&$ -0.079 \pm 0.010 $ & $ -0.037 \pm 0.011$\\
\hline	      	     		     		   
$0.25-0.32$ & $ 0.10 - 0.30 $&$ -0.007 \pm 0.009 $ & $ -0.012 \pm 0.009$\\
      & $ 0.30 - 0.50 $&$ -0.027 \pm 0.008 $ & $ -0.017 \pm 0.008$\\
      & $ 0.50 - 0.64 $&$ -0.053 \pm 0.011 $ & $ -0.029 \pm 0.011$\\
      & $ 0.64 - 1.00 $&$ -0.077 \pm 0.010 $ & $ -0.050 \pm 0.011$\\
\hline	      	     		     		   
$0.32-0.40$ & $ 0.10 - 0.30 $&$ -0.026 \pm 0.011 $ & $  0.019 \pm 0.012$\\ 
      & $ 0.30 - 0.50 $&$ -0.024 \pm 0.010 $ & $ -0.001 \pm 0.010$\\
      & $ 0.50 - 0.64 $&$ -0.055 \pm 0.013 $ & $ -0.055 \pm 0.014$\\
      & $ 0.64 - 1.00 $&$ -0.114 \pm 0.011 $ & $ -0.084 \pm 0.012$\\
\hline	      	     		     		   
$0.40-0.55$ & $ 0.10 - 0.30 $&$ -0.008 \pm 0.011 $ & $  0.006 \pm 0.013$\\ 
      & $ 0.30 - 0.50 $&$ -0.038 \pm 0.010 $ & $ -0.026 \pm 0.011$\\
      & $ 0.50 - 0.64 $&$ -0.066 \pm 0.013 $ & $ -0.078 \pm 0.014$\\
      & $ 0.64 - 1.00 $&$ -0.124 \pm 0.011 $ & $ -0.086 \pm 0.011$\\
\hline	      	     		     		   
$0.55-0.70$ & $ 0.10 - 0.30 $&$ -0.085 \pm 0.018 $ & $ -0.072 \pm 0.020$\\
      & $ 0.30 - 0.50 $&$ -0.110 \pm 0.015 $ & $ -0.109 \pm 0.017$\\
      & $ 0.50 - 0.64 $&$ -0.107 \pm 0.019 $ & $ -0.073 \pm 0.022$\\
      & $ 0.64 - 1.00 $&$ -0.121 \pm 0.015 $ & $ -0.082 \pm 0.018$\\
\hline	      	     		     		   
$0.70-0.85$ & $ 0.10 - 0.30 $&$ -0.176 \pm 0.026 $ & $ -0.202 \pm 0.028$\\
      & $ 0.30 - 0.50 $&$ -0.180 \pm 0.022 $ & $ -0.164 \pm 0.025$\\
      & $ 0.50 - 0.64 $&$ -0.037 \pm 0.029 $ & $ -0.088 \pm 0.033$\\
      & $ 0.64 - 1.00 $&$ -0.076 \pm 0.023 $ & $ -0.029 \pm 0.027$\\

\hline
\end{tabular}
\end{center}
\end{table*}

\begin{table*}
\caption{$A^{UU}_{\cos \phi_h}$ 3d asymmetries asymmetries for positive and negative hadrons in the last $x$
bin (0.038$< x <$0.130).
The errors are statistical only.}
\label{tab:values_3d_c_4}
\begin{center}
\begin{tabular}{ c c r r  }
\hline
$z$ range  &$p_T^{\,h}$ range {\small (GeV/$c$)} & $A^{UU}_{\cos \phi_h}$(h$^+$) & $A^{UU}_{\cos \phi_h}$(h$^-$) \\
\hline
$0.20-0.25$ & $ 0.10 - 0.30$ &$ -0.012 \pm 0.008$ &$  -0.012 \pm 0.009$\\
      & $ 0.30 - 0.50$ &$ -0.031 \pm 0.008$ &$  -0.026 \pm 0.009$\\
      & $ 0.50 - 0.64$ &$ -0.060 \pm 0.011$ &$  -0.022 \pm 0.013$\\
      & $ 0.64 - 1.00$ &$ -0.065 \pm 0.012$ &$  -0.057 \pm 0.012$\\
\hline	      	     		               
$0.25-0.32$ & $ 0.10 - 0.30$ &$ -0.036 \pm 0.009$ &$  -0.026 \pm 0.010$\\
      & $ 0.30 - 0.50$ &$ -0.037 \pm 0.008$ &$  -0.042 \pm 0.009$\\
      & $ 0.50 - 0.64$ &$ -0.052 \pm 0.012$ &$  -0.051 \pm 0.013$\\
      & $ 0.64 - 1.00$ &$ -0.107 \pm 0.011$ &$  -0.057 \pm 0.012$\\
\hline	      	     		               
$0.32-0.40$ & $ 0.10 - 0.30$ &$ -0.035 \pm 0.011$ &$  -0.031 \pm 0.013$\\
      & $ 0.30 - 0.50$ &$ -0.054 \pm 0.010$ &$  -0.019 \pm 0.011$\\
      & $ 0.50 - 0.64$ &$ -0.081 \pm 0.014$ &$  -0.045 \pm 0.015$\\
      & $ 0.64 - 1.00$ &$ -0.094 \pm 0.013$ &$  -0.084 \pm 0.014$\\
\hline	      	     		               
$0.40-0.55$ & $ 0.10 - 0.30$ &$ -0.027 \pm 0.013$ &$  -0.030 \pm 0.014$\\
      & $ 0.30 - 0.50$ &$ -0.042 \pm 0.011$ &$  -0.022 \pm 0.013$\\
      & $ 0.50 - 0.64$ &$ -0.065 \pm 0.014$ &$  -0.071 \pm 0.016$\\
      & $ 0.64 - 1.00$ &$ -0.120 \pm 0.012$ &$  -0.075 \pm 0.013$\\
\hline	      	     		               
$0.55-0.70$ & $ 0.10 - 0.30$ &$ -0.061 \pm 0.021$ &$  -0.068 \pm 0.024$\\
      & $ 0.30 - 0.50$ &$ -0.078 \pm 0.016$ &$  -0.070 \pm 0.020$\\
      & $ 0.50 - 0.64$ &$ -0.116 \pm 0.021$ &$  -0.047 \pm 0.026$\\
      & $ 0.64 - 1.00$ &$ -0.137 \pm 0.016$ &$  -0.058 \pm 0.021$\\
\hline	      	     		               
$0.70-0.85$ & $ 0.10 - 0.30$ &$ -0.129 \pm 0.028$ &$  -0.115 \pm 0.037$\\
      & $ 0.30 - 0.50$ &$ -0.085 \pm 0.026$ &$  -0.159 \pm 0.030$\\
      & $ 0.50 - 0.64$ &$ -0.024 \pm 0.032$ &$  -0.077 \pm 0.044$\\
      & $ 0.64 - 1.00$ &$ -0.039 \pm 0.024$ &$  -0.019 \pm 0.030$\\

\hline
\end{tabular}
\end{center}
\end{table*}

\begin{table*}[htb]
\caption{$A^{UU}_{\cos 2 \phi_h}$ 3d asymmetries asymmetries for positive and negative hadrons in the first $x$
bin (0.003$< x <$0.012).
The errors are statistical only.}
\label{tab:values_3d_c2_1}
\begin{center}
\begin{tabular}{ c c r r  }
\hline
$z$ range & $p_T^{\,h}$ range {\small (GeV/$c$)} & $A^{UU}_{\cos 2\phi_h}$(h$^+$) &  $A^{UU}_{\cos 2\phi_h}$(h$^-$) \\
\hline
$0.20-0.25$ & $ 0.10 - 0.30 $ & $ 0.039 \pm 0.020$ &$ 0.044 \pm 0.021 $ \\ 
      & $ 0.30 - 0.50 $ & $ 0.049 \pm 0.018$ &$ -0.010 \pm 0.020 $  \\
      & $ 0.50 - 0.64 $ & $ 0.013 \pm 0.026$ &$ 0.042 \pm 0.026 $  \\
      & $ 0.64 - 1.00 $ & $ 0.019 \pm 0.025$ &$ 0.058 \pm 0.026 $  \\
\hline	                                   
$0.25-0.32$ & $ 0.10 - 0.30 $ & $ 0.010 \pm 0.022$ &$ -0.005 \pm 0.024 $  \\
      & $ 0.30 - 0.50 $ & $ 0.073 \pm 0.020$ &$ 0.063 \pm 0.021 $  \\
      & $ 0.50 - 0.64 $ & $ 0.018 \pm 0.028$ &$ 0.077 \pm 0.029 $  \\
      & $ 0.64 - 1.00 $ & $-0.004 \pm 0.026$ &$ 0.024 \pm 0.026 $  \\
\hline	                                   
$0.32-0.40$ & $ 0.10 - 0.30 $ & $-0.030 \pm 0.029$ &$ -0.002 \pm 0.030 $  \\
      & $ 0.30 - 0.50 $ & $ 0.102 \pm 0.024$ &$ 0.083 \pm 0.025 $  \\
      & $ 0.50 - 0.64 $ & $-0.007 \pm 0.033$ &$ 0.113 \pm 0.033 $  \\
      & $ 0.64 - 1.00 $ & $ 0.043 \pm 0.029$ &$ 0.112 \pm 0.029 $  \\
\hline	                                   
$0.40-0.55$ & $ 0.10 - 0.30 $ & $ 0.053 \pm 0.031$ &$ 0.142 \pm 0.031 $  \\
      & $ 0.30 - 0.50 $ & $ 0.188 \pm 0.025$ &$ 0.174 \pm 0.026 $  \\
      & $ 0.50 - 0.64 $ & $ 0.025 \pm 0.034$ &$ 0.071 \pm 0.035 $  \\
      & $ 0.64 - 1.00 $ & $ 0.019 \pm 0.029$ &$ 0.039 \pm 0.030 $  \\
\hline	                                   
$0.55-0.70$ & $ 0.10 - 0.30 $ & $ 0.225 \pm 0.047$ &$ 0.213 \pm 0.047 $  \\
      & $ 0.30 - 0.50 $ & $ 0.238 \pm 0.039$ &$ 0.328 \pm 0.038 $  \\
      & $ 0.50 - 0.64 $ & $ 0.132 \pm 0.052$ &$ 0.105 \pm 0.056 $  \\
      & $ 0.64 - 1.00 $ & $-0.134 \pm 0.044$ &$ 0.108 \pm 0.043 $  \\
\hline	                                   
$0.70-0.85$ & $ 0.10 - 0.30 $ & $ 0.174 \pm 0.067$ &$ 0.167 \pm 0.070 $  \\
      & $ 0.30 - 0.50 $ & $ 0.098 \pm 0.061$ &$ 0.170 \pm 0.056 $  \\
      & $ 0.50 - 0.64 $ & $ 0.021 \pm 0.082$ &$ 0.075 \pm 0.088 $  \\
      & $ 0.64 - 1.00 $ & $-0.182 \pm 0.066$ &$ -0.021 \pm 0.067 $  \\

\hline
\end{tabular}
\end{center}
\end{table*}

\begin{table*}[htb]
\caption{$A^{UU}_{\cos 2 \phi_h}$ 3d asymmetries asymmetries for positive and negative hadrons in the second $x$
bin (0.012$<x<$0.020).
The errors are statistical only.}
\label{tab:values_3d_c2_2}
\begin{center}
\begin{tabular}{ c c r r  }
\hline
$z$ range  & $p_T^{\,h}$ range {\small (GeV/$c$)} &  $A^{UU}_{\cos 2\phi_h}$(h$^+$) & $A^{UU}_{\cos 2\phi_h}$(h$^-$) \\
\hline
$0.20-0.25$ & $ 0.10 - 0.30 $ &$ 0.009 \pm 0.015$ &$ -0.013 \pm 0.016 $ \\ 
      & $ 0.30 - 0.50 $ &$-0.003 \pm 0.014$ &$ 0.037 \pm 0.015 $ \\
      & $ 0.50 - 0.64 $ &$ 0.007 \pm 0.020$ &$ 0.009 \pm 0.021 $ \\
      & $ 0.64 - 1.00 $ &$ 0.025 \pm 0.020$ &$ -0.014 \pm 0.021 $ \\
\hline	                                   
$0.25-0.32$ & $ 0.10 - 0.30 $ &$ 0.002 \pm 0.017$ &$ 0.026 \pm 0.018 $ \\
      & $ 0.30 - 0.50 $ &$ 0.013 \pm 0.015$ &$ 0.000 \pm 0.016 $ \\
      & $ 0.50 - 0.64 $ &$ 0.002 \pm 0.021$ &$ 0.005 \pm 0.022 $ \\
      & $ 0.64 - 1.00 $ &$ 0.057 \pm 0.020$ &$ 0.059 \pm 0.021 $ \\
\hline	                                   
$0.32-0.40$ & $ 0.10 - 0.30 $ &$ 0.056 \pm 0.021$ &$ 0.079 \pm 0.023 $ \\
      & $ 0.30 - 0.50 $ &$ 0.054 \pm 0.018$ &$ 0.101 \pm 0.020 $ \\
      & $ 0.50 - 0.64 $ &$ 0.028 \pm 0.024$ &$ 0.012 \pm 0.028 $ \\
      & $ 0.64 - 1.00 $ &$ 0.025 \pm 0.022$ &$ 0.044 \pm 0.024 $ \\
\hline	                                   
$0.40-0.55$ & $ 0.10 - 0.30 $ &$ 0.050 \pm 0.022$ &$ 0.058 \pm 0.025 $ \\
      & $ 0.30 - 0.50 $ &$ 0.090 \pm 0.019$ &$ 0.143 \pm 0.020 $ \\
      & $ 0.50 - 0.64 $ &$ 0.059 \pm 0.026$ &$ 0.091 \pm 0.027 $ \\
      & $ 0.64 - 1.00 $ &$ 0.008 \pm 0.022$ &$ 0.078 \pm 0.023 $ \\
\hline	                                   
$0.55-0.70$ & $ 0.10 - 0.30 $ &$ 0.117 \pm 0.035$ &$ 0.132 \pm 0.036 $ \\
      & $ 0.30 - 0.50 $ &$ 0.158 \pm 0.029$ &$ 0.200 \pm 0.030 $ \\
      & $ 0.50 - 0.64 $ &$ 0.174 \pm 0.037$ &$ 0.064 \pm 0.043 $ \\
      & $ 0.64 - 1.00 $ &$-0.078 \pm 0.032$ &$ 0.052 \pm 0.034 $ \\
\hline	                                   
$0.70-0.85$ & $ 0.10 - 0.30 $ &$ 0.138 \pm 0.048$ &$ 0.143 \pm 0.051 $ \\
      & $ 0.30 - 0.50 $ &$ 0.113 \pm 0.041$ &$ 0.155 \pm 0.045 $ \\
      & $ 0.50 - 0.64 $ &$-0.113 \pm 0.057$ &$ -0.100 \pm 0.071 $ \\ 
      & $ 0.64 - 1.00 $ &$-0.174 \pm 0.049$ &$ 0.020 \pm 0.054 $ \\

\hline
\end{tabular}
\end{center}
\end{table*}

\begin{table*}[htb]
\caption{$A^{UU}_{\cos 2 \phi_h}$ 3d asymmetries asymmetries for positive and negative hadrons in the third $x$
bin (0.020$<x<$0.038).The errors are statistical only.}
\label{tab:values_3d_c2_3}
\begin{center}
\begin{tabular}{ |c |c |r |r | }
\hline
\hline
$z$ range  & $p_T^{\,h}$ range {\small (GeV/$c$)} & $A^{UU}_{\cos 2\phi_h}$(h$^+$) & $A^{UU}_{\cos 2\phi_h}$(h$^-$) \\
\hline
$0.20-0.25$ & $ 0.10 - 0.30 $ &$-0.020 \pm 0.016$ & $-0.030 \pm 0.018 $ \\ 
      & $ 0.30 - 0.50 $ &$-0.003 \pm 0.014$ & $ 0.029 \pm 0.015 $ \\
      & $ 0.50 - 0.64 $ &$ 0.003 \pm 0.021$ & $ 0.038 \pm 0.023 $ \\
      & $ 0.64 - 1.00 $ &$ 0.011 \pm 0.022$ & $ 0.064 \pm 0.023 $ \\
\hline	                                  
$0.25-0.32$ & $ 0.10 - 0.30 $ &$ 0.032 \pm 0.017$ & $ 0.028 \pm 0.018 $ \\
      & $ 0.30 - 0.50 $ &$ 0.028 \pm 0.015$ & $ 0.020 \pm 0.017 $ \\
      & $ 0.50 - 0.64 $ &$-0.005 \pm 0.022$ & $ 0.058 \pm 0.024 $ \\
      & $ 0.64 - 1.00 $ &$ 0.019 \pm 0.021$ & $ 0.052 \pm 0.022 $ \\
\hline	                                  
$0.32-0.40$ & $ 0.10 - 0.30 $ &$ 0.014 \pm 0.022$ & $ 0.091 \pm 0.023 $ \\
      & $ 0.30 - 0.50 $ &$ 0.045 \pm 0.019$ & $ 0.050 \pm 0.021 $ \\
      & $ 0.50 - 0.64 $ &$ 0.016 \pm 0.026$ & $ 0.059 \pm 0.027 $ \\
      & $ 0.64 - 1.00 $ &$ 0.040 \pm 0.023$ & $ 0.075 \pm 0.025 $ \\
\hline	                                  
$0.40-0.55$ & $ 0.10 - 0.30 $ &$ 0.070 \pm 0.022$ & $ 0.066 \pm 0.025 $ \\
      & $ 0.30 - 0.50 $ &$ 0.051 \pm 0.019$ & $ 0.082 \pm 0.022 $ \\
      & $ 0.50 - 0.64 $ &$ 0.097 \pm 0.025$ & $ 0.077 \pm 0.028 $ \\
      & $ 0.64 - 1.00 $ &$-0.014 \pm 0.022$ & $ 0.020 \pm 0.023 $ \\
\hline	                                  
$0.55-0.70$ & $ 0.10 - 0.30 $ &$ 0.090 \pm 0.037$ & $ 0.093 \pm 0.040 $ \\
      & $ 0.30 - 0.50 $ &$ 0.090 \pm 0.029$ & $ 0.129 \pm 0.032 $ \\
      & $ 0.50 - 0.64 $ &$ 0.008 \pm 0.037$ & $ 0.072 \pm 0.044 $ \\
      & $ 0.64 - 1.00 $ &$-0.013 \pm 0.032$ & $ 0.017 \pm 0.038 $ \\
\hline	                                  
$0.70-0.85$ & $ 0.10 - 0.30 $ &$ 0.160 \pm 0.048$ & $ 0.070 \pm 0.054 $ \\
      & $ 0.30 - 0.50 $ &$ 0.074 \pm 0.044$ & $ 0.052 \pm 0.053 $ \\
      & $ 0.50 - 0.64 $ &$-0.038 \pm 0.056$ & $ 0.003 \pm 0.069 $ \\
      & $ 0.64 - 1.00 $ &$-0.077 \pm 0.048$ & $-0.120 \pm 0.057 $ \\
\hline
\hline
\end{tabular}
\end{center}
\end{table*}

\begin{table*}[htb]
\caption{$A^{UU}_{\cos \phi_h}$ 3d asymmetries asymmetries for positive and
 negative hadrons in the last $x$ bin (0.038$< x <$0.130). The errors are
 statistical only.}
\begin{center}
\begin{tabular}{ c c r r  }
\hline
$z$ range  & $p_T^{\,h}$ range {\small (GeV/$c$)} & $A^{UU}_{\cos2 \phi_h}$(h$^+$) &  $A^{UU}_{\cos2 \phi_h}$(h$^-$) \\
\hline
$0.20-0.25$ & $ 0.10 - 0.30 $ &$-0.008 \pm 0.017 $& $ 0.001 \pm 0.018 $ \\ 
      & $ 0.30 - 0.50 $ &$ 0.010 \pm 0.016 $ &$ 0.003 \pm 0.018 $   \\
      & $ 0.50 - 0.64 $ &$ 0.014 \pm 0.023 $ &$ 0.026 \pm 0.026 $   \\
      & $ 0.64 - 1.00 $ &$ 0.074 \pm 0.025 $ &$ 0.041 \pm 0.026 $   \\
\hline	                                   
$0.25-0.32$ & $ 0.10 - 0.30 $ &$-0.004 \pm 0.019 $& $ -0.004 \pm 0.021 $   \\
      & $ 0.30 - 0.50 $ &$-0.021 \pm 0.017 $& $ -0.002 \pm 0.019 $   \\
      & $ 0.50 - 0.64 $ &$-0.005 \pm 0.024 $& $ -0.043 \pm 0.026 $   \\
      & $ 0.64 - 1.00 $ &$ 0.021 \pm 0.023 $ &$ 0.085 \pm 0.024 $   \\
\hline	                                   
$0.32-0.40$ & $ 0.10 - 0.30 $ &$-0.007 \pm 0.024 $& $ 0.003 \pm 0.026 $   \\
      & $ 0.30 - 0.50 $ &$-0.003 \pm 0.020 $& $ 0.036 \pm 0.023 $   \\
      & $ 0.50 - 0.64 $ &$ 0.016 \pm 0.028 $ &$ 0.064 \pm 0.029 $   \\
      & $ 0.64 - 1.00 $ &$ 0.061 \pm 0.025 $ &$ 0.036 \pm 0.028 $   \\
\hline	                                   
$0.40-0.55$ & $ 0.10 - 0.30 $ &$ 0.052 \pm 0.025 $ &$ 0.029 \pm 0.030 $   \\
      & $ 0.30 - 0.50 $ &$ 0.047 \pm 0.021 $ &$ 0.038 \pm 0.025 $   \\
      & $ 0.50 - 0.64 $ &$-0.034 \pm 0.029 $& $ 0.063 \pm 0.033 $   \\
      & $ 0.64 - 1.00 $ &$ 0.018 \pm 0.023 $ &$ 0.091 \pm 0.026 $   \\
\hline	                                   
$0.55-0.70$ & $ 0.10 - 0.30 $ &$ 0.101 \pm 0.040 $ &$ -0.002 \pm 0.048 $   \\
      & $ 0.30 - 0.50 $ &$-0.025 \pm 0.033 $& $ 0.018 \pm 0.041 $   \\
      & $ 0.50 - 0.64 $ &$ 0.046 \pm 0.041 $ &$ 0.033 \pm 0.051 $   \\
      & $ 0.64 - 1.00 $ &$-0.087 \pm 0.034 $& $ 0.027 \pm 0.043 $   \\
\hline	                                   
$0.70-0.85$ & $ 0.10 - 0.30 $ &$-0.060 \pm 0.064 $& $ -0.074 \pm 0.081 $   \\
      & $ 0.30 - 0.50 $ &$-0.004 \pm 0.051 $& $ 0.043 \pm 0.065 $   \\
      & $ 0.50 - 0.64 $ &$-0.043 \pm 0.063 $& $ 0.002 \pm 0.088 $   \\
      & $ 0.64 - 1.00 $ &$-0.068 \pm 0.050 $& $ -0.062 \pm 0.065 $   \\

\hline
\end{tabular}
\end{center}
\label{tab:values_3d_c2_4}
\end{table*}

\FloatBarrier

\bibliographystyle{elsarticle-num}
\bibliography{<your-bib-database>}

\end{document}